\documentclass[%
reprint,
superscriptaddress,
amsmath,
amssymb,
aps,
pra,
longbibliography
]{revtex4-1}

\usepackage{graphicx}
\usepackage{dcolumn}
\usepackage{bm}
\usepackage[T1]{fontenc}
\usepackage[utf8]{inputenc}
\usepackage{times}
\usepackage{amsmath,amsfonts,amssymb,mathtools}
\usepackage{hyperref}
\hypersetup{hypertex=true,
            colorlinks=true,
            linkcolor=red,
            anchorcolor=blue,
            citecolor=blue,
          unicode=true, breaklinks=true}
\usepackage{bbold}
\usepackage{tensor}

\begin{document}

\preprint{APS/123-QED}

\title{Lopsided optical diffraction in loop electromagnetically induced grating}%
\author{Da Huo}
 \affiliation{Center of Quantum Sciences and School of Physics, Northeast Normal University, Changchun 130024, P. R.  China.}
\author{Shuo Hua}
 \affiliation{Center of Quantum Sciences and School of Physics, Northeast Normal University, Changchun 130024, P. R.  China.}
\affiliation{School of Physics, Beihua University, Jilin 132013, P. R.  China.}
\author{Xue-Dong Tian}
\email{snowtxd@gxnu.edu.cn}
\affiliation{College of Physics Science and Technology, Guangxi Normal University, Guilin
541004, P. R. China}
\author{Yi-Mou Liu}%
 \email{liuym605@nenu.edu.cn}
 \affiliation{Center of Quantum Sciences and School of Physics, Northeast Normal University, Changchun 130024, P. R. China.}


\date{\today}

\begin{abstract}
We propose a theoretical scheme in a cold Rubidium-87 ($^{87}$Rb) atomic ensemble with a non-Hermitian optical structure, in which a lopsided optical diffraction grating can be realized just with the combination of single spatially periodic modulation and loop-phase.~Parity-time ($\mathcal{PT}$) symmetric and parity-time antisymmetric ($\mathcal{APT}$) modulation can be switched by adjusting different relative phases of the applied beams. Both $\mathcal{PT}$ symmetry and $\mathcal{PT}$ antisymmetry in our system are robust to the amplitudes of coupling fields,  which allows optical response to be modulated precisely without symmetry breaking. Our scheme shows some nontrivial optical properties, such as lopsided diffraction, single-order diffraction, asymmetric Dammam-like diffraction, etc. Our work will benefit the development of versatile non-Hermitian/asymmetric optical devices.
\end{abstract}

\maketitle


\section{Introduction}

A system with non-Hermitian Hamiltonian being commutative with the parity-time operator ($[\mathcal{PT}, H]=0$),~proposed by Bander \textit{et al}, has a real eigenenergy spectrum and some novel properties under certain conditions \cite{HH1}. Due to the similarity between Schrodinger's equation and the optical Helmholtz equation, the optical system with out-of-phase spatial modulation is a good platform to simulate the system with $\mathcal{PT}$ symmetry, which is named the non-Hermitian optical system \cite{nHOS1}. Corresponding to the potential condition $V(x)=V^{*}(-x)$ satisfied in the $\mathcal{PT}$-symmetry system, the non-Hermitian optical system needs to satisfy complex refractive index condition $n(x)=n^{*}(-x)$ \cite{PT1, PT2, PT3, PT4}. Similarly, optical $\mathcal{PT}$-antisymmetry is realized in optical media with instead condition $n(x)=-n^{*}(-x)$ \cite{APT1,APT2,APT3,APT4,APT5}. In recent years, non-Hermitian optical structures based on discrete systems such as optical waveguide \cite{OW1, OW2}, hybrid optical micro-cavity \cite{MC2, MC3}, electrical circuit resonators \cite{ECR} as well as continuous optical media such as cold atomic ensemble with spatially periodic modulation \cite{PT3, PT4, APT1, APT2, APT3, APT4}, have been implemented successively both in experimental and theoretical works. Moreover, series of new applications or properties have been reported as optical Bloch oscillation \cite{BO1, BO2}, photon or phonon laser \cite{MC1, PL1, PL2}, unidirectional/bidirectional invisibility/optical cloaks \cite{NR1, UI1, UI2, UI3, UI4, UI5, UI6}, and so on. 

The spectroscopic device, such as diffraction grating, has been a significant branch of optical devices since Newton's era.~As an important tool for spectral analysis and optical imaging, it has been playing an important role in many fields such as physics, chemistry, astronomy, biology, etc., \cite{Grating1, Grating2, Grating3, Grating4}.~Electromagnetically induced grating (EIG) \cite{EIG1, EIG2, EIG3} based on electromagnetically induced transparency (EIT) \cite{EIT-review}, makes it possible to tune the diffraction patterns dynamically. Hybrid modulation EIG schemes with traditional amplitude/phase modulations and untraditional ones (nonlinear or nonlocal modulation) have improved diffraction efficiency while greatly expanding the versatility of the grating \cite{EIG4, EIGp1, EIGp2, EIGp3, RydbergEIG1, RydbergEIG2}.  In recent years, combined with the non-Hermitian optical modulation, many schemes of one-/two-dimensional asymmetric optical diffraction gratings have been proposed serially \cite{PTG1, PTG2, PTG3, PTG4, PTG5, PTG6, PTG7, PTG8}, and even similar applications of asymmetric scattering have appeared in the field of acoustics \cite{AA1}. 

However, due to rigid realization conditions of $\mathcal{PT}$/$\mathcal{APT}$ symmetry, there is still a great hindrance to the realization of precise and flexible dynamic operation, especially for some special optical diffractions.~In most previous schemes, dual spatial periodic modulation (via amplitude, detuning of coupling field or atomic density) has been used to achieve two goals, including the realization of $\mathcal{PT}$/$\mathcal{APT}$ symmetry and construction of a grating structure. It results in the lack of accurate modulation capabilities with the protection of $\mathcal{PT}$/$\mathcal{APT}$ symmetry.  Therefore, a method for the preparation of non-Hermitian EIG with simple structure (easy to analyze), dynamic control ability, and protection of optical non-Hermitian symmetry is extremely necessary and desired.

In this paper, we propose a theoretical scheme with a four-level loop-$\mathcal{N}$ configuration in a cold $^{87}$Rb atomic ensemble, in which an EIG with $\mathcal{PT}$-symmetric/-antisymmetric structure can be realized via a single spatially periodic detuning modulation.~$\mathcal{PT}$-symmetric and $\mathcal{PT}$-antisymmetric structures can be easily switched by adjusting different relative phases of the applied beams. In addition, the non-Hermitian optical symmetry in our system is robust to variation of coupling amplitude,  giving our scheme the ability of dynamic modulation with symmetry protection. Considering the contribution of higher-order scattering, which violates Friedel's law \cite{FL1, FL2, FL3}, we also discuss the realization condition of asymmetric diffraction. The perfect lopsided diffraction attained here is explained as the cooperative result of higher-order scattering and spatial Kramers-Kronig relation \cite{SKK1, SKK2, SKK3, SKK4}.  Moreover, we can also achieve single-order and Dammann-like asymmetric diffraction with our scheme. 

This work is organized through the following Sec.~\ref{SecII}, where
we describe the background model, and Sec.~\ref{SecIII}, where we discuss the robustness of optical non-Hermitian symmetry, far-field Fraunhofer lopsided diffraction arising from the combination of higher-order scattering, and spatial Kramers-Kronig relation also with discussions on special asymmetric diffractions based on our scheme.  We summarize, at last, our conclusions in Sec.~\ref{SecIV}.

\begin{figure*}[ptb]
\includegraphics[width=1.0\textwidth]{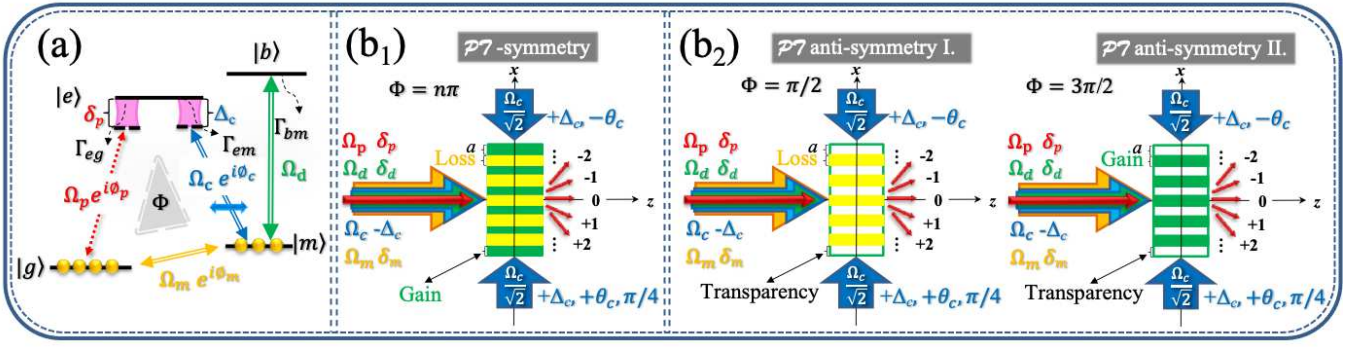}
\caption{A four-level loop-$\mathcal{N}$ configuration (a) for cold $^{87}$Rb atomic Quasi one-dimensional thin grating driven by a probe ($\Omega_p$) field, two coupling ($\Omega_c$ and $\Omega_d$) fields and a microwave ($\Omega_m$) field in $z$ direction, (b$_1$-b$_2$) also with coupling SW components $\Omega_{c1}=\Omega_{c2}=\frac{\Omega_c}{\sqrt{2}}$ along $x$ direction (with small angle $\pm\theta_c$) to give a spatial modulation of detuning $\Delta_c\propto\Delta_{c0}\cdot\sin[C\cdot\pi x]$ as in Ref.~\cite{APT2}. The corresponding initial phases $\phi_{\mu}$ of fields $\Omega_{\mu}$ ($\mu\in\{p,c,m\}$) constitute the loop-phase $\Phi=\phi_m+\phi_c-\phi_p$. The system satisfies $\mathcal{PT}$ symmetry (b$_1$) and $\mathcal{PT}$ antisymmetry (b$_2$), with $\Phi=n\pi$ and $\Phi=\pi/2,3\pi/2$, respectively, so the non-Hermitian symmetry can be switched by controlling the loop-phase $\Phi$.}
\label{Fig1}
\end{figure*}

\section{Model and Equations}\label{SecII}

We start by considering an ensemble of cold $^{87}$Rb atoms driven into a four-level loop-$\mathcal{N}$ configuration, by four coherent fields with frequencies (amplitudes) $\omega_{p}$ ($\mathcal{E}_{p}$), $\omega_{c}$ ($\mathcal{E}_{c}$), $\omega_{d}$ ($\mathcal{E}_{d}$) and $\omega_{m}$ ($\mathcal{E}_{m}$) as shown in Fig.~\ref{Fig1}(a). The weak probe field $\omega_{p}$ interacts with transition $|g\rangle\leftrightarrow|e\rangle$, while the strong control fields $\omega_{c}$ and $\omega_{d}$ act upon transitions $|m\rangle\leftrightarrow|e\rangle$ and $|e\rangle\leftrightarrow|b\rangle$, respectively. The two ground states $|g\rangle$ and $|m\rangle$ are interacted by an equivalent microwave field with $\omega_{m}$. The half Rabi frequencies are defined as $\Omega_{p} =\mathcal{E}_{p} \cdot\wp_{ge}/2\hbar$, $\Omega_{c} = \mathcal{E}_{c}\cdot\wp_{me}/2\hbar$, and $\Omega_{d} = \mathcal{E}_{d}\cdot\wp_{mb}/2\hbar$ with $\omega_{ij}$ being transition frequencies and $\wp_{ij}$ being dipole moments.
In the rotating-wave and electric-dipole approximations, the Hamiltonian for our loop-$\mathcal{N}$-type cold atoms can be written down as $\mathcal{H}=\mathcal{H}_{a}+\mathcal{V}_{af}$ containing an unperturbed atomic Hamiltonian $\mathcal{H}_{a}$ and an atom-field interaction Hamiltonian $\mathcal{V}_{af}$:
\begin{align}
\mathcal{H}_{a} &  =-\hbar\sum^{N}_{j}[\delta_{p}\sigma_{ee}^{j}+(\delta_{p}-\Delta
_{c})\sigma_{mm}^{j} +(\delta_{p}-\Delta_{c}+\delta_{d})\sigma_{bb}^{j}],\nonumber\\
\mathcal{V}_{af} & =-\hbar\sum^{N}_{j}[\Omega_{p}\sigma_{eg}^{j}+\Omega_{c}\sigma_{em}^{j}+\Omega_{d}\sigma_{bm}^{j}+\Omega_{m}\sigma_{mg}^{j}+H.c.],\nonumber\\
\label{Eq_H1}
\end{align}
with $\sigma_{\mu\nu}$ being the projection ($\mu=\nu$) or transition ($\mu\ne\nu$) operators, $\delta_p=\omega_p-\omega_{eg} $ ($\Delta_c=\omega_c-\omega_{em}$, $\delta_d=\omega_d-\omega_{mr}$) is the probe (coupling) detuning.
 Considering the initial phases $\phi_p$, $\phi_c$ and $\phi_m$ of the probe, coupling and microwave fields, we rewrite half Rabi frequencies as $\Omega_p=\widetilde{\Omega}_pe^{i\phi_p}$, $\Omega_c=\widetilde{\Omega}_ce^{i\phi_c}$ and $\Omega_m=\widetilde{\Omega}_me^{i\phi_m}$, where $\widetilde{\Omega}_p$, $\widetilde{\Omega}_c$ and $\widetilde{\Omega}_m$ are chosen to be real, and obtain the density matrix equations (DMEs):
\begin{align*}
\partial_{t}\varrho_{gg} & = \Gamma_{eg}\varrho_{ee}+i\widetilde{\Omega}_p(\varrho_{eg}-\varrho_{ge})+i\widetilde{\Omega}_m(e^{i\Phi}\varrho_{mg}\nonumber\\ &-e^{-i\Phi}\varrho_{gm}),\nonumber\\
\partial_{t}\varrho_{ee} & =
-\Gamma_{em}\varrho_{ee}-\Gamma_{eg}\varrho_{ee}+i\widetilde{\Omega}_p(\varrho_{ge}-\varrho_{eg})+i\widetilde{\Omega}_c(\varrho_{em}\nonumber\\ &-\varrho_{me}),\nonumber\\
\partial_{t}\varrho_{bb} & = -\Gamma_{bm}\varrho_{bb}+ i\widetilde{\Omega}_d(\varrho_{bm}-\varrho_{mb}),\nonumber\\
\partial_{t}\varrho_{gm} & = [\gamma_{gm}+i(\delta_{p}-\Delta_{c})]\varrho_{gm}+ i\widetilde{\Omega}_c\varrho_{ge}-i\widetilde{\Omega}_p\varrho_{em} \nonumber\\ &+i\widetilde{\Omega}_d\varrho_{gb}+i\widetilde{\Omega}_m e^{i\Phi}(\varrho_{gg}-\varrho_{mm}),\nonumber\\
\partial_{t}\varrho_{ge} & = [\gamma_{ge}+i\delta_{p}]\varrho_{ge}+ i\widetilde{\Omega}_c\varrho_{gm}(z)-i\widetilde{\Omega}_me^{i\Phi}\varrho_{me}\nonumber\\ &+i\widetilde{\Omega}_p (\varrho_{gg}-\varrho_{ee}),\nonumber\\
\partial_{t}\varrho_{gb} & = [\gamma_{gb}+i(\delta_{p}-\Delta_{c}+\delta_{d})]\varrho_{gb}+ i\widetilde{\Omega}_d\varrho_{gm}\nonumber\\ &-i\widetilde{\Omega}_m e^{i\Phi}\varrho_{mb}+i\widetilde{\Omega}_p \varrho_{eb},\nonumber\\
\end{align*}
\begin{align}
\partial_{t}\varrho_{me} & = [\gamma_{me}+i\Delta_{c}]\varrho_{me}+i\widetilde{\Omega}_p\varrho_{mg}-i\widetilde{\Omega}_d\varrho_{be}-i\widetilde{\Omega}_m\varrho_{ge}\nonumber\\ &+i\widetilde{\Omega}_c (\varrho_{mm}-\varrho_{ee}),\nonumber\\
\partial_{t}\varrho_{mb} & = [\gamma_{mb}+i\delta_{d}]\varrho_{mb}-i\widetilde{\Omega}_c\varrho_{eb}-i\widetilde{\Omega}_me^{-i\Phi}\varrho_{gb}\nonumber\\ &+i\widetilde{\Omega}_d (\varrho_{mm}-\varrho_{bb}),\nonumber\\
\partial_{t}\varrho_{eb} & = [\gamma_{eb}+i(\delta_{d}-i\Delta_{c})]\varrho_{eb}- i\widetilde{\Omega}_c\varrho_{mb}-i\widetilde{\Omega}_p \varrho_{gb}\nonumber\\ &+i\widetilde{\Omega}_d \varrho_{em},
\label{Eq_DMEs}
\end{align}
with $\varrho_{ij}  = \varrho_{ji}^{*}$. Here $\Phi=\phi_m+\phi_c-\phi_p$ is the relative loop-phase of the three fields.

Setting the time derivatives as zero and utilizing the perturbation method (for $\varepsilon\Omega_p$), the steady-state solution can be attained by solving Eq.~$(\ref{Eq_DMEs})$ analytically:

\begin{align}\label{Eq_rho}
\varrho_{eg} & =\varrho_{eg}^{(0)}+\varepsilon\varrho_{eg}^{(1)}+\varepsilon^2\varrho_{eg}^{(2)}...+\varepsilon^n\varrho_{eg}^{(n)}+...,\\
\varrho_{eg}^{(0)} & =  \frac{\Omega_m\Omega_c}{R_{eg}+\frac{\Omega_{c}^{2}}{R_{mg}+\frac{\Omega_{d}^{2}}{R_{bg}}}}\frac{R_{bg}[\varrho_{mm}^{(0)}-\varrho_{gg}^{(0)}]}{R_{mg}R_{eg}R_{bg}+R_{bg}\Omega_{c}^{2}+R_{eg}\Omega_{d}^{2}},\nonumber\\
\varrho_{eg}^{(1)} & = \frac{e^{-i\Phi}\Omega_{p}}{R_{eg}+\frac{\Omega_{c}^{2}}{R_{mg}+\frac{\Omega_{d}^{2}}{R_{bg}}}}[\frac{(R_{eg}+R_{mg})\Omega_{c}}{R_{mg}R_{eg}}
\cdot\varrho_{mm}^{(0)}-i\varrho_{gg}^{(0)}],\nonumber
\end{align}
with $R_{\mu\nu}=\gamma_{\mu\nu}+i(\omega_{\mu}-\omega_{\nu})$ being the effective coherent relaxation rate between state $|\mu\rangle$ and state $|\nu\rangle$ ($\{\mu,\nu\}=\{g,e,m,b\}$). With condition $\Omega_p\ll\Omega_{c,m,d}$ being satisfied, it is reasonable to take the approximation $\varrho_{eg}\simeq \varrho_{eg}^{(0)}+\varrho_{eg}^{(1)}(\Omega_p\varepsilon)$, where we consider perturbation $\Omega_p\to\Omega_p\varepsilon$ and just keep to the first-order.

The linear probe susceptibility satisfies $\chi_{p}(\omega)=\frac{N\wp_{ge}^2}{2\hbar\varepsilon_{0}}\varrho_{ge}(\omega)/\Omega_{p}$.~$\textbf{Re}[\chi_{p}]$ and $\textbf{Im}[\chi_{p}]$ are used to represent  the real and imaginary parts of probe susceptibility, describing the dispersion and absorption/gain ($\textbf{Im}[\chi_{p}]>0$/$\textbf{Im}[\chi_{p}]<0$) properties, respectively.

Aiming to implement a $nontrivial$ EIG with asymmetric diffraction patterns, $\Delta_{c}$ needs to be periodically modulated as:
\begin{align}
\Delta_{c}(x)=\Delta_{c0}\cdot \sin[\frac{\pi\lambda_{p}(x-x_{0})}{a}].
\label{Eq_delta_c}
\end{align}
For a medium of thickness $\mathcal{L}$ along the $z$ direction and modulated in the $x$ direction, the transmission function of a probe beam takes the form
\begin{align}
T_{\mathcal{\mathcal{L}}}(x)=T_{a}(x)\cdot T_{p}(x),
\label{Eq_T}
\end{align}
where $T_{a}(x)=e^{-k_p\textbf{Im}[\chi_{p}(x)]\mathcal{L}}$ ($T_{p}(x)=e^{ik_p\textbf{Re}[\chi_{p}(x)]\mathcal{L}}$) denotes the amplitude (phase) component with $k_p=2\pi/\lambda_{p}$ being the probe wave vector and $\lambda_{p}$ being the probe wavelength. The Fourier transformation of $T_{\mathcal{L}}(x)$ then yields the Fraunhofer or far-field intensity diffraction equation
\begin{align}
I_{p}(\theta_n)=\frac{{\vert}\mathcal{E}_{p}^{I}(\theta_n){\vert}^{2}\sin^{2}(M{\pi}R\sin\theta_n)}{M^{2}\sin^{2}({\pi}R\sin\theta_n)},
\label{Eq_Ip}
\end{align}
with
\begin{align}
\mathcal{E}_{p}^{I}(\theta_n)=\int_{-a/2}^{+a/2}{T_{\mathcal{L}}(x)e^{-i2{\pi}xR\sin{\theta_n}}}\mathrm{d}x,
\label{Eq_Ep}
\end{align}
with $R=a/{\lambda_{p}}$. In addition, $\theta_n$ denotes the $n$th order diffraction angle of probe photons with respect to the $z$ direction while $M$ represents the ratio between the beam width $\varpi_B$ and the grating period $a$ ($M=\varpi_B/a$). The $n$th-order diffracted probe field will be found at an angle determined by $n=R\sin{\theta_n}\in{(\ldots,-1,0,+1,\ldots)}$.

\section{Results and Discussion}\label{SecIII}

In this section, to implement analytical/numerical calculations based on the above equations, the four atomic levels in Fig.~\ref{Fig1}(a) are assumed as $|g\rangle\equiv|5S_{1/2},F=2\rangle$ and $|m\rangle\equiv|5S_{1/2},F=1\rangle$, as well as $|e\rangle\equiv|5P_{1/2},F=1\rangle$ and $|b\rangle\equiv|5P_{3/2},F=0\rangle$ for cold $^{87}$Rb atoms. Then the decay rates are $\Gamma_{eg}=\Gamma_{em}=\Gamma_{dm}\simeq 3.0\times 2{\pi}$ MHz, with the dephasing rates $\gamma_{eg}=\gamma_{em}\simeq 3.0\times 2{\pi}$ MHz, $\gamma_{md}=\gamma_{gd}=\Gamma_{dm}/2$, and $\gamma_{mg}\simeq 1.0$ kHz. In addition, the atomic medium length is $\mathcal{L}=20.0$ $\mu$m and the density is $N=4.0\times 10^{10}$cm$^{-3}$.

\begin{figure}[ptb]
\includegraphics[width=0.48\textwidth]{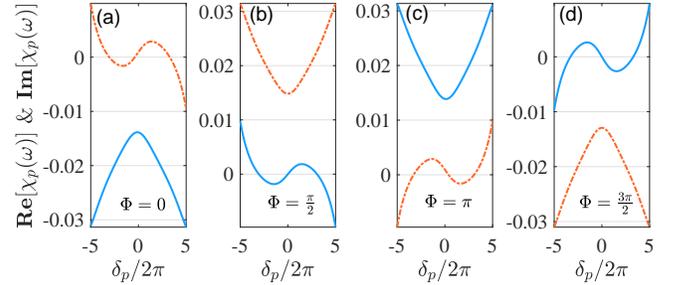}
\caption{Dispersion $\textbf{Re}[\chi_{p}(\omega)]$ (blue-solid curves) and absorption $\textbf{Im}[\chi_{p}(\omega)]$ (orange-dotted curves) as functions of probe detuning $\delta_p$ with $\Delta_{c}=\delta_d=0$ for (a) $\Phi=0$, (b) $\Phi={\pi}$, (c) $\Phi={\pi}/2$, (d) $\Phi=3{\pi}/2$, respectively. Other parameters are chosen as $\Omega_{p}/2\pi=0.01$ MHz, $\Omega_{c}/2\pi=\Omega_{d}/2\pi=5.0$ MHz, and $\Omega_{m}/2\pi=0.5$ MHz.}
\label{Figdp}
\end{figure}

\subsection{$\mathcal{PT}$-symmetry/-antisymmetry based on single periodic modulation}

In the beginning, we apply two control fields ($\Omega_c=\Omega_d=5.0\times 2\pi$ MHz) and an equivalent microwave field ($\Omega_m=0.5\times 2\pi$ MHz), which are all traveling waves (TW). Figure~\ref{Figdp} shows the absorption ($\textbf{Im}[\chi_p(\omega)]$) and dispersion ($\textbf{Re}[\chi_p(\omega)]$) spectra of the system, with detunings $\Delta_c=\delta_d=0$ and different loop phases ($\Phi=0,\pi/2,\pi$ and $3\pi/2$). Evidently, $\textbf{Im}[\chi_p(\omega)]$ ($\textbf{Re}[\chi_p(\omega)]$) is an \textit{odd} (\textit{even}) function of frequency $\omega$ ($\delta_p$) with loop phase $\Phi=n\pi$ as shown in Fig.~\ref{Figdp}(a) and (c). While the parity of $\textbf{Im}[\chi_p(\omega)]$ and $\textbf{Re}[\chi_p(\omega)]$ are opposite if the loop phase is chosen as $\Phi=(2n-1)\pi/2$ in Fig.~\ref{Figdp}(b) and (d), with $n\in \mathbb{Z}$. 

\begin{figure}[ptb]
\includegraphics[width=0.48\textwidth]{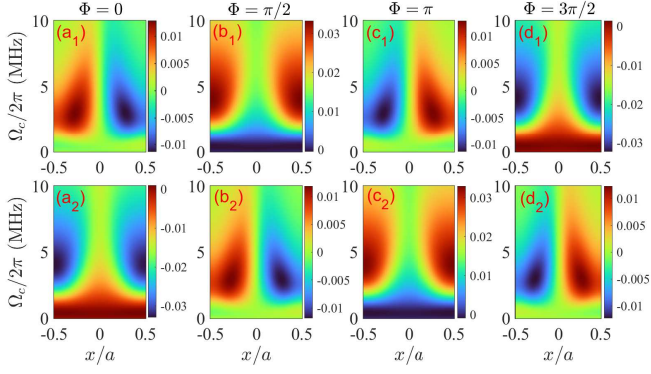}
\caption{Absorption $\textbf{Im}[\chi_{p}(x)])$ (a$_1$-d$_1$) and dispersion $\textbf{Re}[\chi_{p}(x)]$ (a$_2$-d$_2$) versus $x$ and the amplitude of coupling field $\Omega_c$ for different loop phases $\Phi=0, \pi/2, \pi, 3\pi/2$, respectively, and other parameters are same as in Fig.~\ref{Figdp} except $\Delta_{c}=\Delta_{c0}\sin[{\pi}{\lambda_{p}}(x-x_{0})/a]$ and $\Delta_{c0}/2\pi=4.0$ MHz.}
\label{Fig3x}
\end{figure}

Figure~\ref{Fig3x} shows absorption and dispersion spectra via $x$ coordinate when we apply $\Omega_c$ as a standing wave (SW) coupling field, instead of a TW coupling field, with different loop phases $\Phi$. Here we choose the detuning of coupling field  as in Eq.~\eqref{Eq_delta_c} with $\Delta_{c0}=4.0\times 2\pi$ MHz.~Under the two-photon resonance condition ($\delta_d=0$), it can be attained that $\delta_p(x)=-\Delta_c(x)=\Delta_{c0}\cdot\sin[\frac{\pi\lambda_p(x-x_0)}{a}]$. Then it is easy to get $\chi_p(\omega)=\chi_p[\delta_p(x)]\to\chi_p(x)$, indicating the parity characteristic transfer from frequency domain to spatial domain [See Appendix~\ref{S1}].~It is obvious that $\textbf{Im}[\chi_{p}(x)]$ and $\textbf{Re}[\chi_{p}(x)]$ are \textit{odd} (\textit{even}) and \textit{even} (\textit{odd}) functions of $x$ in Fig.~\ref{Fig3x}(a$_1$)-(d$_1$) with $\Phi=0,\pi$ ($\Phi=\pi/2,3\pi/2$), respectively, so the system satisfies optical $\mathcal{PT}$-symmetry ($\mathcal{PT}$-antisymmetry).

\textit{Switching} between $\mathcal{PT}$-symmetric modulation and $\mathcal{PT}$-antisymmetric modulation by adjusting the loop-phase is convenient in our scheme [See Fig.~\ref{Fig3x}, Eq.~\eqref{A2}, and Table.~\ref{tab1}]. With
the loop phase being chosen as $\Phi=2m\pi$ or $\Phi=(2m+1)\pi$ ($m\in \mathbb{Z}$), the system is under $\mathcal{PT}$-symmetric modulation. The real (imaginary) parts $\textbf{Re}[\chi_{p}(x)]$ ($\textbf{Im}[\chi_{p}(x)]$) of the susceptibilities in these two cases with different phases just have the opposite signs. It is worth noting that except loss case (\textit{normal} $\mathcal{APT}$) with $\Phi=\frac{(4m+1)\pi}{2}$, $\mathcal{PT}$ antisymmetric cases also include gain case (\textit{abnormal} $\mathcal{APT}$) with $\Phi=\frac{(4m-1)\pi}{2}$, which is uncommon in the continuous medium schemes.~The gain $\mathcal{APT}$ structures will provide an alternative scheme with high flexibility for some potential applications which require both considerable gain and out-of-phase modulation. 

The parity of the system susceptibility under the loop-phase control makes it possible to achieve both optical $\mathcal{PT}$-symmetry and $\mathcal{PT}$-antisymmetry based on single spatial periodic modulation solely, which is quite different from the previous schemes \cite{PT3, PT4, APT1, APT2, APT3, APT4, PTG1, PTG2, PTG3, PTG4, PTG5, PTG6, PTG7, PTG8}. Especially for $\mathcal{PT}$-symmetry in these continuous periodic systems, the dual spatially periodic modulation, generally provided by multiple SW fields or atomic lattice systems (spatial density modulation), is a necessary condition. Notably, single spatially periodic modulation dramatically reduces the system's complexity and makes it possible to analytically analyze the relationship between non-Hermitian optical symmetry ($\mathcal{PT}$-symmetry/-antisymmetry) and asymmetric diffraction in a system with continuously varying complex susceptibility.

\begin{figure}[ptb]
\includegraphics[width=0.45\textwidth]{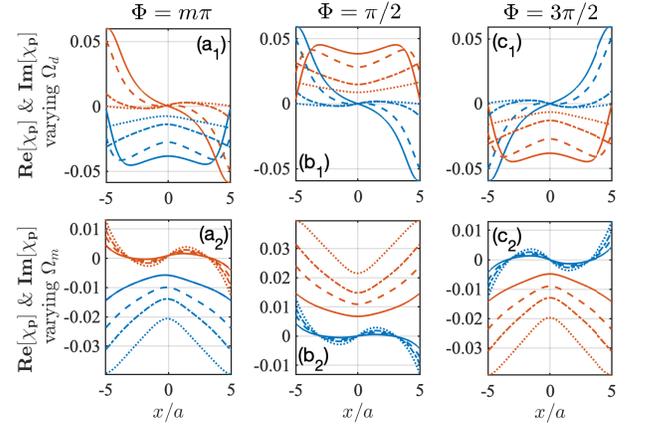}
\caption{Dispersion $\textbf{Re}[\chi_{p}(x)]$ (blue curves) and absorption $\textbf{Im}[\chi_{p}(x)]$ (orange curves) spectra versus $x$ for (a$_1$) $\Phi=0$, (b$_1$) $\Phi={\pi}$, (c$_1$) $\Phi=3{\pi}/2$, respectively, with the same parameters as in Fig.~\ref{Figdp} except $\Omega_d/2\pi=$ 1.0 MHz (solid curves), 2.5 MHz (dashed curves), 5.0 MHz (dashed and dotted curves), and 7.5 MHz (dotted curves).~Moreover, panels (a$_2$)-(c$_2$) are the corresponding spectra with the same parameters as in Fig.~\ref{Figdp} except $\Omega_m/2\pi=$ 0.2 MHz (solid curves), 0.35 MHz (dashed curves), 0.5 MHz (dashed and dotted curves), and 0.8 MHz (dotted curves).}
\label{Fig_odom}
\end{figure}

\begin{table*}[htbp]
\centering
\caption{\it non-Hermitian \bf Spatial Modulation}
\begin{tabular}{cccc}
\hline\hline
 Symmetry Type & Loop-phase $\{\Phi\}$ & Asymmetric Diffraction spectra  & Gain/Loss\\
\hline\hline
$\mathcal{PT}$  & $\{ m\pi\}$, $m\in$ $\mathbb{Z}$ &  \textit{near Lopsided} ($\Omega_c>2\gamma_0$) & \textit{Both} (\bf{balance})\\
\textit{normal} $\mathcal{APT}$ & $\{(4m+1)\pi/2\}$ & \textit{Perfect Lopsided}  ($\Omega_c>2\gamma_0$) &  \textit{Loss} (\bf{pure})\\
\textit{abnormal} $\mathcal{APT}$ & $\{(4m-1)\pi/2\}$ & \textit{Irregular}  ($\Omega_c>2\gamma_0$) &  \textit{Gain} (\bf{pure})\\
\hline\hline
\end{tabular}
  \label{tab1}
\end{table*}

\textit{Robustness} of non-Hermitian optical symmetry is necessary to be considered for a system that requires precise control, as the symmetry needs to be satisfied strictly to implement some special optical functions (such as perfect lopsided diffraction) [See Sec.~\ref{B}]. In previous works, the non-Hermitian symmetries are very fragile to even a slight perturbation of parameters such as coupling amplitudes. Here we consider the robustness of our system with varying coupling intensity $\Omega_c$ in Fig.~\ref{Fig3x}. It is obvious that the system has a large dynamic tuning range here ($\Omega_c$ $\in$ $[2.2,10.0]\times 2\pi$ MHz), with the protection of non-Hermitian optical symmetry. Similarly, non-Hermitian symmetry robustness to other varying coupling amplitudes ($\Omega_d$ and $\Omega_m$) are shown in Fig.~\ref{Fig_odom}. The above conclusion can also be analytically supported by Eq.~\eqref{A1}. Compared to the optical depth (OD) modulation (via initial preparation of atomic density $N$ and medium length $\mathcal{L}$) in the previous work \cite{PTG2}, the amplitude modulation in our scheme is more feasible with a large dynamic modulation range under non-Hermitian optical symmetry protecting, meaning the possibility to achieve some special optical functions.

 \subsection{Lopsided, Single-order, and Dammann-like asymmetric Diffractions \label{B}}
 \label{B}
 
\begin{figure}[ptb]
\includegraphics[width=0.48\textwidth]{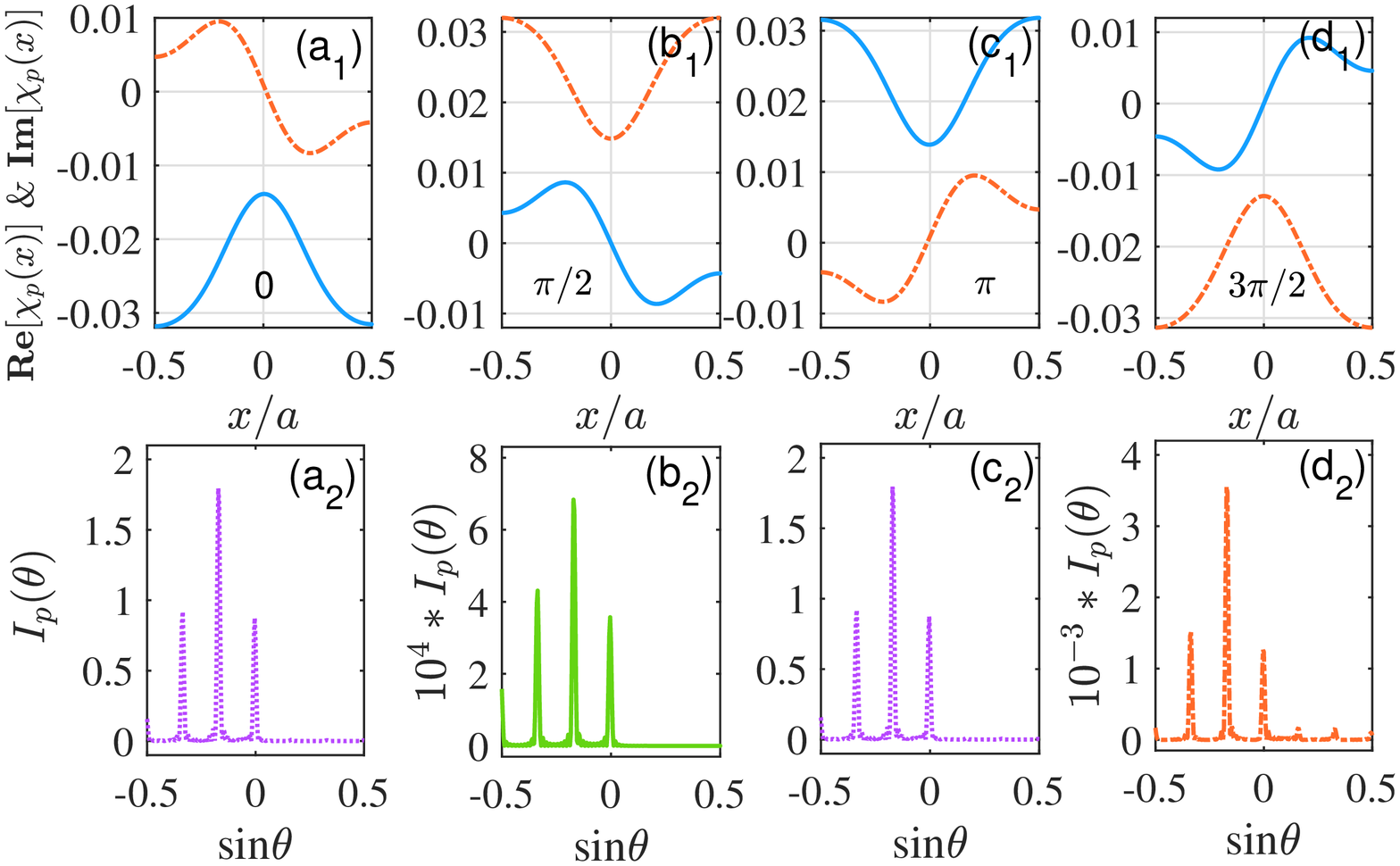}
\caption{Dispersion $\textbf{Re}[\chi_{p}(x)]$ (blue-solid curves) and absorption $\textbf{Im}[\chi_{p}(x)]$ (orange-dotted curves) spectra of gratings with lopsided diffraction patterns versus $x$ with $\Delta_{c}=\Delta_{c0}\sin[{\pi}{\lambda_{p}}(x-x_{0})/a]$ and $\delta_{p}=0$ for (a$_1$) $\Phi=0$, (b$_1$) $\Phi={\pi}$, (c$_1$) $\Phi={\pi}/2$, (d$_1$) $\Phi=3{\pi}/2$, respectively. Moreover, (a$_2$)-(d$_2$) are the corresponding grating diffraction intensity angle spectrum with the same parameters as in Fig.~\ref{Figdp} except $\Delta_{c0}/2\pi=4.0$ MHz, $R=6$ and $M=10$.}
\label{Figxa}
\end{figure}

After constructing the tunable optical $\mathcal{PT}$/$\mathcal{APT}$ symmetry of the system with the loop phase, we try to analyze the diffraction characteristics of our system in this subsection.~The optical diffraction characteristics of this \textit{non-Hermitian} EIG can be studied by injecting a probe field along the $z$-axis which is perpendicular to the direction of spatial periodic modulation ($x$-axis).~Figs.~\ref{Figxa}(a$_1$-d$_1$) give dispersion and absorption (or gain) spectra under the out-of-phase ($\mathcal{PT}$-symmetric/-antisymmetric) modulations for different loop-phases, with same parameters as in Fig.~\ref{Figdp} except $\Delta_{c0}/2\pi=4.0$ MHz. Figs.~\ref{Figxa}(a$_2$-d$_2$) show that the probe beam is only diffracted into the negative angles in the range $\theta\in(-\pi/6,0)$ with loop phase $\Phi=0,\pi/2,\pi,3\pi/2$, respectively. Fig.~\ref{Figxa}(b$_2$) displays \textit{normal} $\mathcal{APT}$ case, that is, the system is pure dissipative. Instead, Fig.~\ref{Figxa}(d$_2$) exhibits an unconventional optical $\mathcal{PT}$-antisymmetric case with optical gain, which is also quite different from the $\mathcal{PT}$ symmetric case [See Fig.~\ref{Figxa}(a$_2$) and Fig.~\ref{Figxa}(c$_2$)] with the balance of gain and loss.
 
 It is known that diffraction peaks occur at discrete angles $\theta_n$ determined by $k_n=k_p\sin\theta_n=2n\pi/a$ or $n=R\sin\theta_n$, where we choose $R = a/\lambda_p=6$ and $M = \varpi_B/a=10$. Combined with Eq.~\eqref{Eq_T}, we could focus on the $n$th-order diffraction by examining $\mathcal{E}_n=\mathcal{E}_p(\theta_n)$ for $n\neq 0$. For simplicity, with $\alpha(x)=k_p\textbf{Im}[\chi_p(x)]\mathcal{L}$, $\beta(x)=k_p\textbf{Re}[\chi_p(x)]\mathcal{L}$, and $\gamma_n(x)=2n\pi x$, we can make a power series expansion of Eq.~\eqref{Eq_Ep} and obtain

 \begin{align}
 \mathcal{E}^I_n & =\int^{+a/2}	_{-a/2}dx \cdot e^{-i\gamma_n(x)}[1+\alpha(x)+\frac{\alpha(x)^2}{2}+\frac{\alpha(x)^3}{6}...]\nonumber\\
 & \cdot[1+i\beta(x)-\frac{\beta(x)^2}{2}
 -\frac{i\beta(x)^3}{6}+...+\frac{(i\beta_{x})^m}{m!}...],
 \end{align}
where it is easy to find $|\alpha(x)|\sim|\beta(x)|\ll 1$ ($N=4.0\times$10$^{10}$cm$^{-3}$ and $\mathcal{L}=20\mu m$). Defining
 \begin{align}\label{Eq_fg}
 f_n^{\prime} & =\int^{+a/2}_{-a/2}dx\cdot[1+\alpha(x)+\frac{\alpha(x)^2}{2}]\beta(x)\sin[\gamma_n(x)],\\
 f_n^{\prime\prime} & =\int^{+a/2}_{-a/2}dx\cdot[1+\alpha(x)+\frac{\alpha(x)^2}{2}]\beta(x)\cos[\gamma_n(x)],\nonumber\\
 g_n^{\prime} & =\int^{+a/2}_{-a/2}dx\cdot[1+\alpha(x)+\frac{\alpha(x)^2}{2}]\frac{\beta(x)^2-2}{2}\cos[\gamma_n(x)],\nonumber\\
 g_n^{\prime\prime} & =\int^{+a/2}_{-a/2}dx\cdot[1+\alpha(x)+\frac{\alpha(x)^2}{2}]\frac{\beta(x)^2-2}{2}\sin[\gamma_n(x)],\nonumber
 \end{align}
 with the replacement $\beta(x)\to \varepsilon_n\beta(x)$, we further get
 \begin{align}\label{Eq_En}
 \mathcal{E}_n\simeq [f_n^{\prime}\varepsilon_n-g_n^{\prime}\varepsilon_n^2/2]+i[f_n^{\prime\prime}\varepsilon_n+g_n^{\prime\prime}\varepsilon_n^2/2],
 \end{align}
 where the scattering factor $\varepsilon_n$ is small enough to keep only the first- and second-order scattering terms \cite{HS}.
 It is easy to find that $f_n^{\prime}=-f_{-n}^{\prime}$, $f_n^{\prime\prime}=f_{-n}^{\prime\prime}$, $g_n^{\prime}=g_{-n}^{\prime}$ and $g_n^{\prime\prime}=-g_{-n}^{\prime\prime}$, and we can write down the intensities $I_{\pm n}\simeq |f_n^{\prime}\varepsilon_n\mp g_n^{\prime}\varepsilon_n^2/2|^2+|f_n^{\prime\prime}\varepsilon_n\pm g_n^{\prime\prime}\varepsilon_n^2/2|^2$ for the $\pm n$th diffraction orders. Accordingly, the intensity contrast ratio can be introduced as
 \begin{equation}\label{eta}
 \eta_n=\left\vert\frac{I_n-I_{-n}}{I_n+I_{-n}}\right\vert\simeq \left\vert\frac{f_n^{\prime}\cdot g_n^{\prime}-f_n^{\prime\prime}\cdot g_n^{\prime\prime}}{(f_n^{\prime})^2+(f_n^{\prime\prime})^2}\right\vert\cdot\varepsilon_n,	
 \end{equation}
  to evaluate the degree of asymmetric diffraction. 
  
  Considering $\gamma_n(-x)=-\gamma_n(x)$, it is not difficult to find $\eta_n\equiv 0$ ($I_{n}=I_{-n}$) with $f_n^{\prime}\cdot g_n^{\prime}=f_n^{\prime\prime}\cdot g_n^{\prime\prime}$, indicating the symmetric diffraction occurs in the system, in the case of spatial even symmetry with $\textbf{Im}[\chi_p(x)]=\textbf{Im}[\chi_p(-x)]$ and $\textbf{Re}[\chi_p(x)]=\textbf{Re}[\chi_p(-x)]$ (in-phase modulation). On the contrary, if the system is deviated from in-phase modulation, with $f_n^{\prime}\cdot g_n^{\prime}\neq f_n^{\prime\prime}\cdot g_n^{\prime\prime}$ ($\eta_n\neq 0$), we can obtain the asymmetric diffraction angle spectra. Moreover, if optical systems satisfy $\mathcal{PT}$-symmetry or $\mathcal{PT}$-antisymmetry (out-of-phase) there will be a large possibility to achieve $\eta_n\to 1$, named the perfect asymmetric diffraction or lopsided diffraction [See Appendix~\ref{S2}].

\begin{figure}[ptb]
\includegraphics[width=0.5\textwidth]{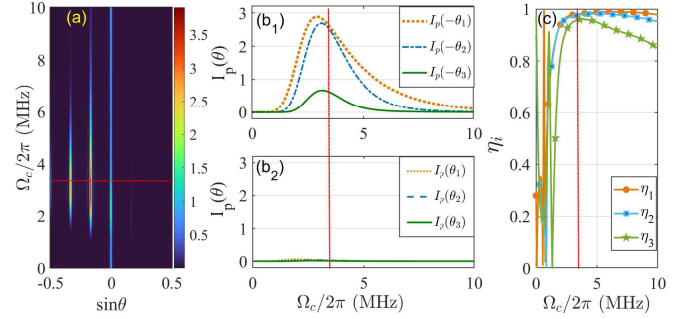}
\caption{Diffraction intensity $I_p(\theta)$ versus the diffraction angle and the amplitude of coupling field $\Omega_c$ (a), negative order diffraction intensity $I_p(-\theta_n)$ (b$_1$), positive order diffraction intensity $I_p(\theta_n)$ (b$_2$) and asymmetric diffraction coefficient $\eta_i$ versus $\Omega_c$ (c), with the same parameters as in Fig.~\ref{Figxa} except $\Phi=0$. The position of the red line corresponds to $\Omega_c/2\pi=3.5$ MHz.}
\label{FigIeta}
\end{figure}

Fig.~\ref{FigIeta}(a) shows the asymmetric diffraction angular spectra $I_p(\theta)$ with varying coupling amplitude $\Omega_c$ of our system for the $\mathcal{PT}$ symmetric case ($\Phi=0$). Panels (b$_1$) and (b$_2$) show the $\pm n$th order diffraction intensity $I_p(\pm\theta_{n})$ ($n=\{1,2,3\}$) varying with the coupling amplitude ($\Omega_c/2\pi\in(0,10)$ MHz). Moreover, we show the intensity contrast ratio (asymmetry diffraction coefficient) $\eta_n$ versus $\Omega_c$ in Fig.~\ref{FigIeta}(c). Comparing the three panels of Fig.~\ref{FigIeta}, it is easy to find that, in the range of $\Omega_c/2\pi\in(2.0,5.0)$ MHz, a high degree of asymmetric diffraction is obtained ($\eta_{n}>0.95$ and $\to$ 1), with one-sided non-zero diffraction intensity ($I_p(-\theta_n)>0.2$). Particularly, we can get $\eta_n>0.97$ and $I_p(\theta_n)>0.6$, which is extremely close to the optimal situation for $\mathcal{PT}$-symmetric case, at the red line position ($\Omega_c/2\pi=3.5$ MHz).

\begin{figure}[ptb]
\includegraphics[width=0.48\textwidth]{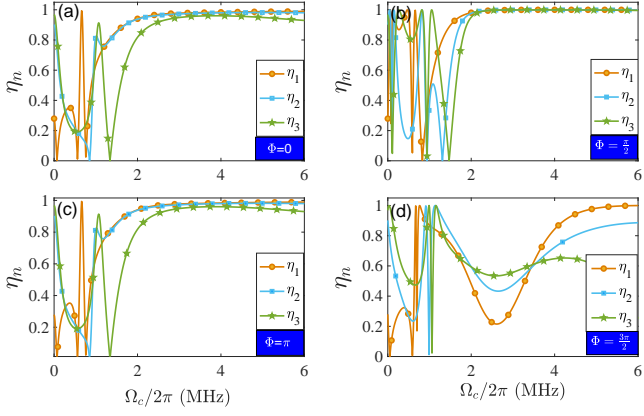}
\caption{Asymmetric diffraction coefficient $\eta_n$ versus $\Omega_c$, $n\in\{1,2,3\}$, with the same parameters as in Fig.~\ref{Figxa} except loop-phase $\Phi=$0, $\pi/2$, $\pi$ and $3\pi/2$ in panel (a), (b), (c) and (d), respectively.}
\label{Fig_eta}
\end{figure}

Next, we further discuss asymmetric diffraction of the system with different loop phases. The curves of asymmetry diffraction coefficients versus $\Omega_c$ for different diffraction orders $\eta_n$ ($n=\{1,2,3\}$) are shown in Fig.~\ref{Fig_eta} with different loop phases $\Phi$.~Comparing the $\mathcal{PT}$ symmetric modulation cases in Fig.~\ref{Fig_eta}(a) and Fig.~\ref{Fig_eta}(c), we can obtain the asymmetry diffraction coefficients $\eta_{1,2}\to 1$ ($0.85<\eta_{3}<0.95$) in the coupling amplitude range $\Omega_c/2\pi>2.0$ MHz, and values of $\eta_{1,2}$ reach the maximum at $\Omega_c/2\pi=3.5$ MHz, so these cases can be named as \textit{near lopsided} diffractions in Table.~\ref{tab1}. 

Figure~\ref{Fig_eta}(b) shows asymmetry diffraction coefficients $\eta_n$ in \textit{normal} $\mathcal{PT}$-antisymmetric case with $\Phi=\pi/2$. Obviously, the \textit{perfect lopsided} diffraction effect has been achieved, that is, $\eta_{n}\equiv 1$ ($n=\{1,2,3\}$) if the coupling amplitude is chosen as $\Omega_c/2\pi>2.0$ MHz. Employing the same parameters as in Fig.~\ref{Figxa} (b), for the pure-loss optical system, we can easily obtain 
\begin{subequations}
\begin{align}
\eta_n^{\mathcal{APT}} & =\left\vert \frac{f_n^{\prime}g_n^{\prime}}{f_n^{\prime}f_n^{\prime}}\right\vert\cdot\varepsilon_n=\left\vert \frac{g_n^{\prime}}{f_n^{\prime}}\right\vert\cdot\varepsilon_n \label{eta_APTa}\\
& \simeq\left\vert \frac{\int_0^{a/2}dx \beta(x)\sin[2n\pi x]}{\int_0^{a/2}dx [1+\alpha(x)]\cos[2n\pi x]}\right\vert\cdot\varepsilon_n \label{eta_APTb}\\
& =\left\vert \frac{\int_0^{a/2}dx \alpha(x)\cos[2n\pi x]}{\int_0^{a/2}dx \alpha(x)\cos[2n\pi x]}\right\vert=1, \label{eta_APTc}
\end{align}
\end{subequations}
with $\varepsilon_n=1$ from Eq.~\eqref{B3}. Here we also use $\beta(x)\propto\int \alpha(x)dx$ ($\alpha(x)\propto\int \beta(x)dx$) from the spatial Kramers-Kronig relations of susceptibility \cite{SKK1, SKK2, SKK3, SKK4} 
\begin{align}
\textbf{Re}[\chi_p(x)] & =\frac{1}{\pi}\textbf{P}\int \frac{\textbf{Im}[\chi_p(x^{\prime})]}{x^{\prime}-x}dx^{\prime},\nonumber\\
\textbf{Im}[\chi_p(x)] & =\frac{1}{\pi}\textbf{P}\int \frac{\textbf{Re}[\chi_p(x^{\prime})]}{x^{\prime}-x}dx^{\prime},
\end{align}
where \textbf{P} indicates the principal value of the integral. \textit{Perfect Lopsided} diffraction is the cooperative result of multiple higher-order scattering, which can be explained by Eq.~\eqref{Eq_En} and Eq.~\eqref{eta}, and the spatial Kramers-Kronig relations. For the gain $\mathcal{APT}$ case, \textit{irregular} asymmetric diffraction is shown in Fig.~\ref{Fig_eta}(d). The perfect lopsided diffraction condition will be broken if the condition of series expansion is not satisfied any longer, owing to the large optical gain ($\alpha(x)\sim\beta(x)\gg 1$). 

\begin{figure}[ptb]
\includegraphics[width=0.48\textwidth]{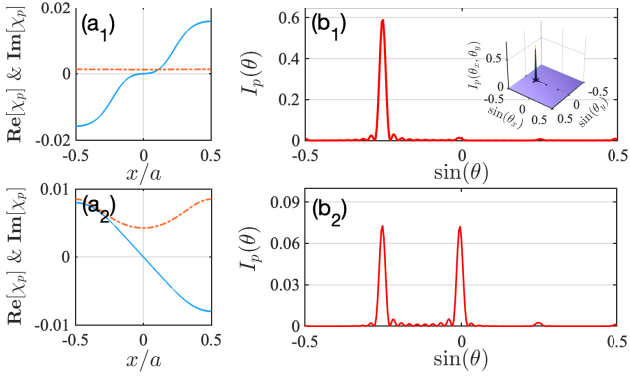}
\caption{\textit{Single-order} (a$_1$-b$_1$) and \textit{Dammann-like} asymmetric (a$_2$-b$_2$) diffraction. Dispersion $\textbf{Re}[\chi_{p}(x)]$ (blue-solid curves)/absorption $\textbf{Im}[\chi_{p}(x)]$ (orange-dotted curves) spectra versus $x$ in (a$_1$-a$_2$) and  diffraction intensity $I_p(\theta)$ versus $\sin(\theta)$ in (b$_1$-b$_2$) for 1D case, with $\Omega_c/2\pi=1.056$ MHz$,\Omega_c/2\pi=3.0$ MHz$,\Omega_c/2\pi=0.7$ MHz for (a$_1$-b$_1$); with the same parameters as in Fig.~\ref{Figxa} except $\Omega_c/2\pi=1.55$ MHz for (a$_2$-b$_2$).  Diffraction intensity $I_p(\theta_x,\theta_y)$ for 2D case in (b$_{1}^{\text{in}}$) with $\Delta_c=\Delta_{cx}\sin[\pi\lambda_p(x-x_0)/a]+\Delta_{cy}\cos[\pi\lambda_p(y-y_0)/a]$, $\Delta_{cx}/2\pi=4.0$ MHz and $\Delta_{cy}/2\pi=0.1$ MHz.}
\label{Fig_SD}
\end{figure}

With the tunable ability of lopsided diffraction in our scheme, we could discuss some special diffractions further. \textit{Single-order diffraction} is displayed in Fig.~\ref{Fig_SD} (a$_1$) and (b$_1$) for the optical $\mathcal{APT}$ case ($\Phi=\pi/2$), with showing the susceptibility and diffraction angle spectrum of the one-dimensional case (1D), respectively. Here the parameters are chosen as $\Omega_c=1.056\gamma_0,\Omega_c=3.0\gamma_0$, and $\Omega_c=0.7\gamma_0$ ($\gamma_0=1.0\times 2\pi$ MHz). It is obvious that the probe beam is only diffracted into the $-1$st order with a considerable diffraction efficiency (intensity) $I_{-1}\simeq$ 0.597. In addition, the above conclusion can be extended to the two-dimensional case (2D). We plot the single-order asymmetric diffraction for 2D case in Fig.~\ref{Fig_SD}(b$_{1}^{\text{in}}$) with spatial periodic modulation in both directions ($\Delta_c=\Delta_{cx}\sin[\pi\lambda_p(x-x_0)/a]+\Delta_{cy}\sin[\pi\lambda_p(y-y_0)/a]$). Here, only a simple case ($\Delta_{cx}/2\pi=4.0$ MHz and $\Delta_{cy}/2\pi=0.1$ MHz) is given to illustrate the functionality of our scheme. In fact, in combination with the two-dimensional Hermitian/non-Hermitian hybrid modulation \cite{PTG7}, there will be other diffraction modes of single-order asymmetric diffraction.

Identically, through continuous parameter modulation for several coupling fields ($\Omega_{c,d,m}$, $\Delta_{c0}$, $\delta_{p,d}$), the special asymmetric diffraction with multiple equal intensity diffraction orders can be easily accomplished. As the simplest example, the case with two equal-intensity diffraction orders is shown in Fig.~\ref{Fig_SD} (a$_2$) and (b$_2$). Here we first check the susceptibilities and diffraction angle spectra in the 1D case to ensure the satisfaction of non-Hermitian optical symmetry and lopsided diffraction conditions. In the same way, following the 2D non-Hermitian hybrid modulation methods \cite{PTG7}, an array of diffraction beams with equal intensity will be implemented, named asymmetric \textit{Dammann-like} diffraction grating, the analog of the Dammann gratings \cite{DAMMANN_1, DAMMANN_2, DAMMANN_3, DAMMANN_4, DAMMANN_5} in the asymmetric case.~These special asymmetric diffractions and the implementation methods will greatly facilitate the development and application of scattering-type asymmetric optical devices.

\section{Conclusions}\label{SecIV}

In summary, the ensemble of cold $^{87}$Rb atoms driven into a four-level loop-$\mathcal{N}$ configuration can provide an interesting venue to realize non-Hermitian EIG with various symmetry features.~Firstly, the scheme is proposed to prepare continuous out-of-phase type non-Hermitian optical media ($\mathcal{PT}$-symmetric/-antisymmetric) based on loop-phase ( $\Phi=\phi_m+\phi_c-\phi_p$) and single spatially periodic modulation. $\mathcal{PT}$ symmetry and $\mathcal{PT}$ antisymmetry can be easily switched by the loop-phase.~Then robustness of non-Hermitian optical symmetry of this scheme to the coupling amplitudes is discussed.~It shows that our system has a flexible multi-parameter modulation ability with a large dynamic adjusting range while protecting the non-Hermitian optical symmetry. In addition, we examine the far-field Fraunhofer diffraction of the atomic grating, including $\mathcal{PT}$ symmetry cases as well as loss $\mathcal{PT}$-antisymmetric case (\textit{normal} $\mathcal{APT}$) and gain $\mathcal{PT}$ antisymmetric (\textit{abnormal} $\mathcal{APT}$) case. Moreover, we analyze the reason for asymmetric diffraction and discuss the realization condition of the perfect lopsided diffraction, interpreted as the joint contribution of higher-order scattering and the spatial Kramers-Kronig relationship of complex susceptibility. Finally, special diffractions, single-order/Dammann-like asymmetric diffraction, are shown to demonstrate the superiority of our scheme.

The introduction of loop-phase modulation reduces the complexity of achieving non-Hermitian optical symmetry in a continuous medium and diversifies the methods of spatial modulation, which will promote and benefit the development of signal-selecting non-Hermitian/asymmetric optical devices.

\textbf{Acknowledgments}

This work is supported by the National Natural Science Foundation of China (Grant~No.~12104107), and the Natural Science Foundation of Guangxi Province (Grants No.~AD19245180).

\appendix

\section{Susceptibility}
\label{S1}
The susceptibility of the system satisfies $\chi_p(x)=\frac{\mu\mathcal{E}_p}{\hbar\varepsilon_0}\cdot\frac{\varrho_{eg}}{\Omega_p}\propto \varrho_{eg}^{(0)}(x)+\varepsilon\varrho_{eg}^{(1)}(x)+...$, where we cut off at first order owning to $\Omega_p\ll \gamma$.~With dephasing $\gamma_{eg}=\gamma_{em}=\gamma_{dm}=\gamma$, $\gamma_{mg}= \gamma_{m}$ ($\gamma_m\ll \gamma$) and the two-photon resonance condition $\delta=\delta_p-\Delta_c(x)=0$, it is obvious that $\delta_p(x)=\Delta_c(x)$ and $\varrho_{eg}(x)$ are the function of $x$. Then, by rewriting the steady-state solutions in Eq.~\eqref{Eq_rho} in terms of real and imaginary parts, we can get  $\varrho_{eg}^{(0)}(x)=\textbf{A}_{0}[\delta_p(x)^{n_{A0}}]-i\textbf{B}_{0}[\delta_p(x)^{n_{B0}}]$ and $\varrho_{eg}^{(1)}(x)=e^{-i\Phi}\textbf{A}_{1}[\delta_p(x)^{n_{A1}}]-ie^{-i\Phi}\textbf{B}_{1}[\delta_p(x)^{n_{B1}}]$. Here $n_{\zeta}$ $(\zeta\in\{A_0,B_0,A_1,B_1\})$ is the highest power of $\delta_p(x)$ in the numerator of the corresponding polynomial, which will determine the parity of the functions if each term of polynomial has the same denominator. We can get
\begin{widetext}
\begin{align}\label{A1}
\textbf{A}_{0}[\delta_p(x)^{n_{A0}}] & \simeq \gamma[\varrho_{mm}^{(0)}-\varrho_{gg}^{(0)}]\frac{\Omega_m\Omega_c}{\Omega_d^2}\cdot \frac{\gamma^2+\delta_p^2}{(\gamma^2+\delta_p^2)^2},n_{A0}\to 2,\nonumber\\
\textbf{B}_{0}[\delta_p(x)^{n_{B0}}] & \simeq \gamma[\varrho_{mm}^{(0)}-\varrho_{gg}^{(0)}]\frac{\Omega_m\Omega_c}{\Omega_d^2}\cdot \frac{2\delta_p}{(\gamma^2+\delta_p^2)^2}, n_{B0}\to 1,\nonumber\\
\textbf{A}_{1}[\delta_p(x)^{n_{A1}}] & \simeq \Omega_c\varrho_{mm}^{(0)}\frac{(\gamma+\gamma_m)(\delta_p^2+\gamma^2)}{\gamma_m(\gamma^2+\delta_p^2)^2}-\Omega_p\varrho_{gg}^{(0)}\frac{\gamma_m(\gamma^2\delta_p+\delta_p^3)}{\gamma_m(\gamma^2+\delta_p^2)^2},n_{A1}\to 3,\nonumber\\
\textbf{B}_{1}[\delta_p(x)^{n_{B1}}] & \simeq \Omega_c\varrho_{mm}^{(0)}\frac{\gamma^2\delta_p}{\gamma_m(\gamma^2+\delta_p^2)^2}+\Omega_p\varrho_{gg}^{(0)}\frac{\gamma_m\gamma(\gamma^2+\delta_p^2)}{\gamma_m(\gamma^2+\delta_p^2)^2},n_{B1}\to 2.
\end{align}
\end{widetext}
 Moreover, based on Eq.~\eqref{Eq_rho} we can easily attain the relationship between real (imaginary) part $\textbf{Re}[\chi_p]$ ($\textbf{Im}[\chi_p]$) of susceptibility and space coordinates $x$:

\begin{widetext}
\begin{align}\label{A2}
\chi_p(x) & \propto \textbf{A}_{0}[\varrho_{eg}^{(0)}(x)] -i\textbf{B}_{0}[\varrho_{eg}^{(0)}(x)] +e^{-i\Phi}\textbf{A}_{1}[\varrho_{eg}^{(1)}(x)] -ie^{-i\Phi}\textbf{B}_{1}[\varrho_{eg}^{(1)}(x)],\\
\textbf{Re}[\chi_p(x)] &\propto \left\{\begin{aligned}
\textbf{A}_{0}[\delta_p(x)^2]\pm\textbf{A}_{1}[\delta_p(x)^3], &&\Phi =n\pi,&&\textbf{Odd}\nonumber\\
\textbf{A}_{0}[\delta_p(x)^2]\pm\textbf{B}_{1}[\delta_p(x)^2], &&\Phi =\frac{(2n-1)\pi}{2},   &&\textbf{Even}\\
\end{aligned}
\right.\\
\textbf{Im}[\chi_p(x)] &\propto \left\{\begin{aligned}
\textbf{B}_{0}[\delta_p(x)^1]\pm\textbf{B}_{1}[\delta_p(x)^2], &&\Phi =n\pi, &&\textbf{Even}\\
\textbf{B}_{0}[\delta_p(x)^1]\mp\textbf{A}_{1}[\delta_p(x)^3], &&\Phi =\frac{(2n-1)\pi}{2}, &&\textbf{Odd}.\nonumber\\
\end{aligned}
\right.
\end{align}
\end{widetext}

The real and imaginary parts of $\varrho_{eg}^{(1)}$ introduced by the loop structure finally determine the parity of susceptibility of the system. Hence the non-Hermitian optical ($\mathcal{PT}$- or $\mathcal{APT}$-) symmetry can be easily modulated by loop-phase $\Phi$. Significantly, from Eq.~\eqref{A1}, we can find the varying coupling amplitudes $\Omega_{\mu}$ ($\mu=\{c,d,m\}$) have no impact on the parity of susceptibility of our system under the reasonable condition.

\section{Asymmetric Diffraction}
\label{S2}
In this part, we will try to calculate  $\eta_n$ according to the spatial symmetry of the system. Firstly, for general Hermitian optical media, the real and imaginary parts of susceptibility are even functions of $x$, $\textbf{Im}[\chi_p(x)]=\textbf{Im}[\chi_p(-x)]$ and $\textbf{Re}[\chi_p(x)]=\textbf{Re}[\chi_p(-x)]$. We can obtain $\alpha(x)=\alpha(-x)$ and $\beta(x)=\beta(-x)$, where $\alpha(x)=k_p\textbf{Im}[\chi_p(x)]\mathcal{L}$ and $\beta(x)=k_p\textbf{Re}[\chi_p(x)]\mathcal{L}$. Combined with the above relationship and Eq.~\eqref{Eq_fg}, the scattering coefficients read:
\begin{widetext}
\begin{align}
f_n^{\prime} & =\int^{a/2}_{0}dx\cdot[1+\alpha(x)+\frac{\alpha(x)^2}{2}]\beta(x)\sin[\gamma_n(x)] -\int^{a/2}_{0}dx\cdot[1+\alpha(x)+\frac{\alpha(x)^2}{2}]\beta(x)\sin[\gamma_n(x)]=0,\nonumber\\
g_n^{\prime\prime} & =\int^{a/2}_{0}dx\cdot[1+\alpha(x)+\frac{\alpha(x)^2}{2}]\frac{\beta(x)^2-2}{2}\sin[\gamma_n(x)] -\int^{a/2}_{0}dx\cdot[1+\alpha(x)+\frac{\alpha(x)^2}{2}]\frac{\beta(x)^2-2}{2}\sin[\gamma_n(x)]=0.\nonumber
\end{align}
\end{widetext}
Apparently, we can get the result $\eta_n=0$ for each $n$th diffraction order, meaning a symmetric diffraction behavior of the Hermitian grating.

In a similar fashion, with non-Hermitian modulations, we can get
\begin{widetext}
\begin{align}\label{B1B2}
[f_n^{\prime}\cdot g_n^{\prime}]_{\mathcal{PT}} & =\int^{a/2}_{0}dx\cdot 2\alpha(x)\beta(x)\sin[\gamma_n(x)] \times\int^{a/2}_{0}dx\cdot[1+\frac{\alpha(x)^2}{2}][\beta(x)^2-2]\cos[\gamma_n(x)],\nonumber\\
[f_n^{\prime\prime}\cdot g_n^{\prime\prime}]_{\mathcal{PT}} & =\int^{a/2}_{0}dx\cdot[1+\frac{\alpha(x)^2}{2}]\beta(x)\sin[\gamma_n(x)] \times\int^{a/2}_{0}dx\cdot\alpha(x)[\beta(x)^2-2]\cos[\gamma_n(x)],\\
[f_n^{\prime}\cdot g_n^{\prime}]_{\mathcal{APT}} & =\int^{a/2}_{0}dx\cdot[2+2\alpha(x)+\alpha(x)^2]\beta(x)\sin[\gamma_n(x)]\times\int^{a/2}_{0}dx\cdot[1+\alpha(x)+\frac{\alpha(x)^2}{2}][\beta(x)^2-2]\cos[\gamma_n(x)],\nonumber\\
[f_n^{\prime\prime}\cdot g_n^{\prime\prime}]_{\mathcal{APT}} & =0,
\end{align}
\end{widetext}
with $\alpha(x)\to$ \textit{odd}, $\beta(x)\to$ \textit{even} ($\mathcal{PT}$-symmetric case) and $\alpha(x)\to$ \textit{even}, $\beta(x)\to$ \textit{odd} ($\mathcal{PT}$-antisymmetric case), respectively.
After simplification, we can attain the relationship between the asymmetry coefficients and the scattering coefficients
\begin{align}\label{B3}
\eta_n^{\mathcal{PT}} & =\left\vert \frac{[f_n^{\prime}\cdot g_n^{\prime}]_{\mathcal{PT}}-[f_n^{\prime\prime}\cdot g_n^{\prime\prime}]_{\mathcal{PT}}}{[f_n^{\prime}]_{\mathcal{PT}}^2+[f_n^{\prime\prime}]_{\mathcal{PT}}^2}\right\vert\cdot\varepsilon_n,\nonumber\\
\eta_n^{\mathcal{APT}} & =\left\vert \frac{[f_n^{\prime}\cdot g_n^{\prime}]_{\mathcal{APT}}}{[f_n^{\prime}]_{\mathcal{APT}}^2+[f_n^{\prime\prime}]_{\mathcal{APT}}^2}\right\vert\cdot\varepsilon_n,
\end{align}
 corresponding to two different non-Hermitian optical symmetries, respectively.

Thus, we know that when the system satisfies non-Hermitian symmetry, $\eta_n^{\mathcal{PT}}$ ($\eta_n^{\mathcal{APT}}$) is unequal to 0, which causes asymmetric diffraction behavior of the atomic grating. Moreover, the contribution of high-order scattering can be influenced by adjusting parameters such as coupling amplitudes. It gives a method to achieve adjusting the degree of asymmetry accurately.

\

\bibliographystyle{apsrev4-1}
\bibliography{Manu_V6.bib}

\begin{thebibliography}{64}%
\makeatletter
\providecommand \@ifxundefined [1]{%
 \@ifx{#1\undefined}
}%
\providecommand \@ifnum [1]{%
 \ifnum #1\expandafter \@firstoftwo
 \else \expandafter \@secondoftwo
 \fi
}%
\providecommand \@ifx [1]{%
 \ifx #1\expandafter \@firstoftwo
 \else \expandafter \@secondoftwo
 \fi
}%
\providecommand \natexlab [1]{#1}%
\providecommand \enquote  [1]{``#1''}%
\providecommand \bibnamefont  [1]{#1}%
\providecommand \bibfnamefont [1]{#1}%
\providecommand \citenamefont [1]{#1}%
\providecommand \href@noop [0]{\@secondoftwo}%
\providecommand \href [0]{\begingroup \@sanitize@url \@href}%
\providecommand \@href[1]{\@@startlink{#1}\@@href}%
\providecommand \@@href[1]{\endgroup#1\@@endlink}%
\providecommand \@sanitize@url [0]{\catcode `\\12\catcode `\$12\catcode
  `\&12\catcode `\#12\catcode `\^12\catcode `\_12\catcode `\%12\relax}%
\providecommand \@@startlink[1]{}%
\providecommand \@@endlink[0]{}%
\providecommand \url  [0]{\begingroup\@sanitize@url \@url }%
\providecommand \@url [1]{\endgroup\@href {#1}{\urlprefix }}%
\providecommand \urlprefix  [0]{URL }%
\providecommand \Eprint [0]{\href }%
\providecommand \doibase [0]{http://dx.doi.org/}%
\providecommand \selectlanguage [0]{\@gobble}%
\providecommand \bibinfo  [0]{\@secondoftwo}%
\providecommand \bibfield  [0]{\@secondoftwo}%
\providecommand \translation [1]{[#1]}%
\providecommand \BibitemOpen [0]{}%
\providecommand \bibitemStop [0]{}%
\providecommand \bibitemNoStop [0]{.\EOS\space}%
\providecommand \EOS [0]{\spacefactor3000\relax}%
\providecommand \BibitemShut  [1]{\csname bibitem#1\endcsname}%
\let\auto@bib@innerbib\@empty
\bibitem [{\citenamefont {Bender}(2007)}]{HH1}%
  \BibitemOpen
  \bibfield  {author} {\bibinfo {author} {\bibfnamefont {C.~M.}\ \bibnamefont
  {Bender}},\ }\href {\doibase 10.1088/0034-4885/70/6/r03} {\bibfield
  {journal} {\bibinfo  {journal} {Rep. Prog. Phys.}\ }\textbf {\bibinfo
  {volume} {70}},\ \bibinfo {pages} {947} (\bibinfo {year} {2007})}\BibitemShut
  {NoStop}%
\bibitem [{\citenamefont {El-Ganainy}\ \emph {et~al.}(2007)\citenamefont
  {El-Ganainy}, \citenamefont {Makris}, \citenamefont {Christodoulides},\ and\
  \citenamefont {Musslimani}}]{nHOS1}%
  \BibitemOpen
  \bibfield  {author} {\bibinfo {author} {\bibfnamefont {R.}~\bibnamefont
  {El-Ganainy}}, \bibinfo {author} {\bibfnamefont {K.~G.}\ \bibnamefont
  {Makris}}, \bibinfo {author} {\bibfnamefont {D.~N.}\ \bibnamefont
  {Christodoulides}}, \ and\ \bibinfo {author} {\bibfnamefont {Z.~H.}\
  \bibnamefont {Musslimani}},\ }\href {\doibase 10.1364/OL.32.002632}
  {\bibfield  {journal} {\bibinfo  {journal} {Opt. Lett.}\ }\textbf {\bibinfo
  {volume} {32}},\ \bibinfo {pages} {2632} (\bibinfo {year}
  {2007})}\BibitemShut {NoStop}%
\bibitem [{\citenamefont {Makris}\ \emph {et~al.}(2008)\citenamefont {Makris},
  \citenamefont {El-Ganainy}, \citenamefont {Christodoulides},\ and\
  \citenamefont {Musslimani}}]{PT1}%
  \BibitemOpen
  \bibfield  {author} {\bibinfo {author} {\bibfnamefont {K.~G.}\ \bibnamefont
  {Makris}}, \bibinfo {author} {\bibfnamefont {R.}~\bibnamefont {El-Ganainy}},
  \bibinfo {author} {\bibfnamefont {D.~N.}\ \bibnamefont {Christodoulides}}, \
  and\ \bibinfo {author} {\bibfnamefont {Z.~H.}\ \bibnamefont {Musslimani}},\
  }\href {\doibase 10.1103/PhysRevLett.100.103904} {\bibfield  {journal}
  {\bibinfo  {journal} {Phys. Rev. Lett.}\ }\textbf {\bibinfo {volume} {100}},\
  \bibinfo {pages} {103904} (\bibinfo {year} {2008})}\BibitemShut {NoStop}%
\bibitem [{\citenamefont {Klaiman}\ \emph {et~al.}(2008)\citenamefont
  {Klaiman}, \citenamefont {G\"unther},\ and\ \citenamefont {Moiseyev}}]{PT2}%
  \BibitemOpen
  \bibfield  {author} {\bibinfo {author} {\bibfnamefont {S.}~\bibnamefont
  {Klaiman}}, \bibinfo {author} {\bibfnamefont {U.}~\bibnamefont {G\"unther}},
  \ and\ \bibinfo {author} {\bibfnamefont {N.}~\bibnamefont {Moiseyev}},\
  }\href {\doibase 10.1103/PhysRevLett.101.080402} {\bibfield  {journal}
  {\bibinfo  {journal} {Phys. Rev. Lett.}\ }\textbf {\bibinfo {volume} {101}},\
  \bibinfo {pages} {080402} (\bibinfo {year} {2008})}\BibitemShut {NoStop}%
\bibitem [{\citenamefont {Hang}\ \emph {et~al.}(2013)\citenamefont {Hang},
  \citenamefont {Huang},\ and\ \citenamefont {Konotop}}]{PT3}%
  \BibitemOpen
  \bibfield  {author} {\bibinfo {author} {\bibfnamefont {C.}~\bibnamefont
  {Hang}}, \bibinfo {author} {\bibfnamefont {G.}~\bibnamefont {Huang}}, \ and\
  \bibinfo {author} {\bibfnamefont {V.~V.}\ \bibnamefont {Konotop}},\ }\href
  {\doibase 10.1103/PhysRevLett.110.083604} {\bibfield  {journal} {\bibinfo
  {journal} {Phys. Rev. Lett.}\ }\textbf {\bibinfo {volume} {110}},\ \bibinfo
  {pages} {083604} (\bibinfo {year} {2013})}\BibitemShut {NoStop}%
\bibitem [{\citenamefont {Sheng}\ \emph {et~al.}(2013)\citenamefont {Sheng},
  \citenamefont {Miri}, \citenamefont {Christodoulides},\ and\ \citenamefont
  {Xiao}}]{PT4}%
  \BibitemOpen
  \bibfield  {author} {\bibinfo {author} {\bibfnamefont {J.}~\bibnamefont
  {Sheng}}, \bibinfo {author} {\bibfnamefont {M.-A.}\ \bibnamefont {Miri}},
  \bibinfo {author} {\bibfnamefont {D.~N.}\ \bibnamefont {Christodoulides}}, \
  and\ \bibinfo {author} {\bibfnamefont {M.}~\bibnamefont {Xiao}},\ }\href
  {\doibase 10.1103/PhysRevA.88.041803} {\bibfield  {journal} {\bibinfo
  {journal} {Phys. Rev. A}\ }\textbf {\bibinfo {volume} {88}},\ \bibinfo
  {pages} {041803} (\bibinfo {year} {2013})}\BibitemShut {NoStop}%
\bibitem [{\citenamefont {Ge}\ and\ \citenamefont {T\"ureci}(2013)}]{APT1}%
  \BibitemOpen
  \bibfield  {author} {\bibinfo {author} {\bibfnamefont {L.}~\bibnamefont
  {Ge}}\ and\ \bibinfo {author} {\bibfnamefont {H.~E.}\ \bibnamefont
  {T\"ureci}},\ }\href {\doibase 10.1103/PhysRevA.88.053810} {\bibfield
  {journal} {\bibinfo  {journal} {Phys. Rev. A}\ }\textbf {\bibinfo {volume}
  {88}},\ \bibinfo {pages} {053810} (\bibinfo {year} {2013})}\BibitemShut
  {NoStop}%
\bibitem [{\citenamefont {Wu}\ \emph {et~al.}(2014)\citenamefont {Wu},
  \citenamefont {Artoni},\ and\ \citenamefont {La~Rocca}}]{APT2}%
  \BibitemOpen
  \bibfield  {author} {\bibinfo {author} {\bibfnamefont {J.-H.}\ \bibnamefont
  {Wu}}, \bibinfo {author} {\bibfnamefont {M.}~\bibnamefont {Artoni}}, \ and\
  \bibinfo {author} {\bibfnamefont {G.~C.}\ \bibnamefont {La~Rocca}},\ }\href
  {\doibase 10.1103/PhysRevLett.113.123004} {\bibfield  {journal} {\bibinfo
  {journal} {Phys. Rev. Lett.}\ }\textbf {\bibinfo {volume} {113}},\ \bibinfo
  {pages} {123004} (\bibinfo {year} {2014})}\BibitemShut {NoStop}%
\bibitem [{\citenamefont {Wu}\ \emph {et~al.}(2015)\citenamefont {Wu},
  \citenamefont {Artoni},\ and\ \citenamefont {La~Rocca}}]{APT3}%
  \BibitemOpen
  \bibfield  {author} {\bibinfo {author} {\bibfnamefont {J.-H.}\ \bibnamefont
  {Wu}}, \bibinfo {author} {\bibfnamefont {M.}~\bibnamefont {Artoni}}, \ and\
  \bibinfo {author} {\bibfnamefont {G.~C.}\ \bibnamefont {La~Rocca}},\ }\href
  {\doibase 10.1103/PhysRevA.91.033811} {\bibfield  {journal} {\bibinfo
  {journal} {Phys. Rev. A}\ }\textbf {\bibinfo {volume} {91}},\ \bibinfo
  {pages} {033811} (\bibinfo {year} {2015})}\BibitemShut {NoStop}%
\bibitem [{\citenamefont {Peng}\ \emph {et~al.}(2016)\citenamefont {Peng},
  \citenamefont {Cao}, \citenamefont {Shen}, \citenamefont {Qu}, \citenamefont
  {Wen}, \citenamefont {Jiang},\ and\ \citenamefont {Xiao}}]{APT4}%
  \BibitemOpen
  \bibfield  {author} {\bibinfo {author} {\bibfnamefont {P.}~\bibnamefont
  {Peng}}, \bibinfo {author} {\bibfnamefont {W.}~\bibnamefont {Cao}}, \bibinfo
  {author} {\bibfnamefont {C.}~\bibnamefont {Shen}}, \bibinfo {author}
  {\bibfnamefont {W.}~\bibnamefont {Qu}}, \bibinfo {author} {\bibfnamefont
  {J.}~\bibnamefont {Wen}}, \bibinfo {author} {\bibfnamefont {L.}~\bibnamefont
  {Jiang}}, \ and\ \bibinfo {author} {\bibfnamefont {Y.}~\bibnamefont {Xiao}},\
  }\href {\doibase 10.1038/nphys3842} {\bibfield  {journal} {\bibinfo
  {journal} {Nature Phys.}\ }\textbf {\bibinfo {volume} {12}},\ \bibinfo
  {pages} {1139} (\bibinfo {year} {2016})}\BibitemShut {NoStop}%
\bibitem [{\citenamefont {Yang}\ \emph {et~al.}(2017)\citenamefont {Yang},
  \citenamefont {Liu},\ and\ \citenamefont {You}}]{APT5}%
  \BibitemOpen
  \bibfield  {author} {\bibinfo {author} {\bibfnamefont {F.}~\bibnamefont
  {Yang}}, \bibinfo {author} {\bibfnamefont {Y.-C.}\ \bibnamefont {Liu}}, \
  and\ \bibinfo {author} {\bibfnamefont {L.}~\bibnamefont {You}},\ }\href
  {\doibase 10.1103/PhysRevA.96.053845} {\bibfield  {journal} {\bibinfo
  {journal} {Phys. Rev. A}\ }\textbf {\bibinfo {volume} {96}},\ \bibinfo
  {pages} {053845} (\bibinfo {year} {2017})}\BibitemShut {NoStop}%
\bibitem [{\citenamefont {Nguyen}\ \emph {et~al.}(2016)\citenamefont {Nguyen},
  \citenamefont {Maier}, \citenamefont {Hong},\ and\ \citenamefont
  {Oulton}}]{OW1}%
  \BibitemOpen
  \bibfield  {author} {\bibinfo {author} {\bibfnamefont {N.~B.}\ \bibnamefont
  {Nguyen}}, \bibinfo {author} {\bibfnamefont {S.~A.}\ \bibnamefont {Maier}},
  \bibinfo {author} {\bibfnamefont {M.}~\bibnamefont {Hong}}, \ and\ \bibinfo
  {author} {\bibfnamefont {R.~F.}\ \bibnamefont {Oulton}},\ }\href {\doibase
  10.1088/1367-2630/18/12/125012} {\bibfield  {journal} {\bibinfo  {journal}
  {New J. Phys.}\ }\textbf {\bibinfo {volume} {18}},\ \bibinfo {pages} {125012}
  (\bibinfo {year} {2016})}\BibitemShut {NoStop}%
\bibitem [{\citenamefont {Eichelkraut}\ \emph {et~al.}(2013)\citenamefont
  {Eichelkraut}, \citenamefont {Heilmann}, \citenamefont {Weimann},
  \citenamefont {Stützer}, \citenamefont {Dreisow}, \citenamefont
  {Christodoulides}, \citenamefont {Nolte},\ and\ \citenamefont
  {Szameit}}]{OW2}%
  \BibitemOpen
  \bibfield  {author} {\bibinfo {author} {\bibfnamefont {T.}~\bibnamefont
  {Eichelkraut}}, \bibinfo {author} {\bibfnamefont {R.}~\bibnamefont
  {Heilmann}}, \bibinfo {author} {\bibfnamefont {S.}~\bibnamefont {Weimann}},
  \bibinfo {author} {\bibfnamefont {S.}~\bibnamefont {Stützer}}, \bibinfo
  {author} {\bibfnamefont {F.}~\bibnamefont {Dreisow}}, \bibinfo {author}
  {\bibfnamefont {D.~N.}\ \bibnamefont {Christodoulides}}, \bibinfo {author}
  {\bibfnamefont {S.}~\bibnamefont {Nolte}}, \ and\ \bibinfo {author}
  {\bibfnamefont {A.}~\bibnamefont {Szameit}},\ }\href {\doibase
  10.1038/ncomms3533} {\bibfield  {journal} {\bibinfo  {journal} {Nat.
  Commun.}\ }\textbf {\bibinfo {volume} {4}},\ \bibinfo {pages} {2533}
  (\bibinfo {year} {2013})}\BibitemShut {NoStop}%
\bibitem [{\citenamefont {Peng}\ \emph {et~al.}(2014)\citenamefont {Peng},
  \citenamefont {Özdemir}, \citenamefont {Lei}, \citenamefont {Monifi},
  \citenamefont {Gianfreda}, \citenamefont {Long}, \citenamefont {Fan},
  \citenamefont {Nori}, \citenamefont {Bender},\ and\ \citenamefont
  {Yang}}]{MC2}%
  \BibitemOpen
  \bibfield  {author} {\bibinfo {author} {\bibfnamefont {B.}~\bibnamefont
  {Peng}}, \bibinfo {author} {\bibfnamefont {S.~K.}\ \bibnamefont {Özdemir}},
  \bibinfo {author} {\bibfnamefont {F.}~\bibnamefont {Lei}}, \bibinfo {author}
  {\bibfnamefont {F.}~\bibnamefont {Monifi}}, \bibinfo {author} {\bibfnamefont
  {M.}~\bibnamefont {Gianfreda}}, \bibinfo {author} {\bibfnamefont {G.~L.}\
  \bibnamefont {Long}}, \bibinfo {author} {\bibfnamefont {S.}~\bibnamefont
  {Fan}}, \bibinfo {author} {\bibfnamefont {F.}~\bibnamefont {Nori}}, \bibinfo
  {author} {\bibfnamefont {C.~M.}\ \bibnamefont {Bender}}, \ and\ \bibinfo
  {author} {\bibfnamefont {L.}~\bibnamefont {Yang}},\ }\href {\doibase
  10.1038/nphys2927} {\bibfield  {journal} {\bibinfo  {journal} {Nature Phys.}\
  }\textbf {\bibinfo {volume} {10}},\ \bibinfo {pages} {394} (\bibinfo {year}
  {2014})}\BibitemShut {NoStop}%
\bibitem [{\citenamefont {Chang}\ \emph {et~al.}(2014)\citenamefont {Chang},
  \citenamefont {Jiang}, \citenamefont {Hua}, \citenamefont {Yang},
  \citenamefont {Wen}, \citenamefont {Jiang}, \citenamefont {Li}, \citenamefont
  {Wang},\ and\ \citenamefont {Xiao}}]{MC3}%
  \BibitemOpen
  \bibfield  {author} {\bibinfo {author} {\bibfnamefont {L.}~\bibnamefont
  {Chang}}, \bibinfo {author} {\bibfnamefont {X.}~\bibnamefont {Jiang}},
  \bibinfo {author} {\bibfnamefont {S.}~\bibnamefont {Hua}}, \bibinfo {author}
  {\bibfnamefont {C.}~\bibnamefont {Yang}}, \bibinfo {author} {\bibfnamefont
  {J.}~\bibnamefont {Wen}}, \bibinfo {author} {\bibfnamefont {L.}~\bibnamefont
  {Jiang}}, \bibinfo {author} {\bibfnamefont {G.}~\bibnamefont {Li}}, \bibinfo
  {author} {\bibfnamefont {G.}~\bibnamefont {Wang}}, \ and\ \bibinfo {author}
  {\bibfnamefont {M.}~\bibnamefont {Xiao}},\ }\href {\doibase
  10.1038/nphoton.2014.133} {\bibfield  {journal} {\bibinfo  {journal} {Nature
  Photon.}\ }\textbf {\bibinfo {volume} {8}},\ \bibinfo {pages} {524} (\bibinfo
  {year} {2014})}\BibitemShut {NoStop}%
\bibitem [{\citenamefont {Choi}\ \emph {et~al.}(2018)\citenamefont {Choi},
  \citenamefont {Hahn}, \citenamefont {Yoon},\ and\ \citenamefont
  {Song}}]{ECR}%
  \BibitemOpen
  \bibfield  {author} {\bibinfo {author} {\bibfnamefont {Y.}~\bibnamefont
  {Choi}}, \bibinfo {author} {\bibfnamefont {C.}~\bibnamefont {Hahn}}, \bibinfo
  {author} {\bibfnamefont {J.~W.}\ \bibnamefont {Yoon}}, \ and\ \bibinfo
  {author} {\bibfnamefont {S.~H.}\ \bibnamefont {Song}},\ }\href {\doibase
  10.1038/s41467-018-04690-y} {\bibfield  {journal} {\bibinfo  {journal} {Nat.
  Commun.}\ }\textbf {\bibinfo {volume} {9}},\ \bibinfo {pages} {2182}
  (\bibinfo {year} {2018})}\BibitemShut {NoStop}%
\bibitem [{\citenamefont {Longhi}(2009)}]{BO1}%
  \BibitemOpen
  \bibfield  {author} {\bibinfo {author} {\bibfnamefont {S.}~\bibnamefont
  {Longhi}},\ }\href {\doibase 10.1103/PhysRevLett.103.123601} {\bibfield
  {journal} {\bibinfo  {journal} {Phys. Rev. Lett.}\ }\textbf {\bibinfo
  {volume} {103}},\ \bibinfo {pages} {123601} (\bibinfo {year}
  {2009})}\BibitemShut {NoStop}%
\bibitem [{\citenamefont {Zhang}\ \emph {et~al.}(2017)\citenamefont {Zhang},
  \citenamefont {Zhang}, \citenamefont {Zhang}, \citenamefont {Li},
  \citenamefont {Zhang}, \citenamefont {Li}, \citenamefont {Beli\'{c}},\ and\
  \citenamefont {Xiao}}]{BO2}%
  \BibitemOpen
  \bibfield  {author} {\bibinfo {author} {\bibfnamefont {Y.}~\bibnamefont
  {Zhang}}, \bibinfo {author} {\bibfnamefont {D.}~\bibnamefont {Zhang}},
  \bibinfo {author} {\bibfnamefont {Z.}~\bibnamefont {Zhang}}, \bibinfo
  {author} {\bibfnamefont {C.}~\bibnamefont {Li}}, \bibinfo {author}
  {\bibfnamefont {Y.}~\bibnamefont {Zhang}}, \bibinfo {author} {\bibfnamefont
  {F.}~\bibnamefont {Li}}, \bibinfo {author} {\bibfnamefont {M.~R.}\
  \bibnamefont {Beli\'{c}}}, \ and\ \bibinfo {author} {\bibfnamefont
  {M.}~\bibnamefont {Xiao}},\ }\href {\doibase 10.1364/OPTICA.4.000571}
  {\bibfield  {journal} {\bibinfo  {journal} {Optica}\ }\textbf {\bibinfo
  {volume} {4}},\ \bibinfo {pages} {571} (\bibinfo {year} {2017})}\BibitemShut
  {NoStop}%
\bibitem [{\citenamefont {Hodaei}\ \emph {et~al.}(2014)\citenamefont {Hodaei},
  \citenamefont {Miri}, \citenamefont {Heinrich}, \citenamefont
  {Christodoulides},\ and\ \citenamefont {Khajavikhan}}]{MC1}%
  \BibitemOpen
  \bibfield  {author} {\bibinfo {author} {\bibfnamefont {H.}~\bibnamefont
  {Hodaei}}, \bibinfo {author} {\bibfnamefont {M.-A.}\ \bibnamefont {Miri}},
  \bibinfo {author} {\bibfnamefont {M.}~\bibnamefont {Heinrich}}, \bibinfo
  {author} {\bibfnamefont {D.~N.}\ \bibnamefont {Christodoulides}}, \ and\
  \bibinfo {author} {\bibfnamefont {M.}~\bibnamefont {Khajavikhan}},\ }\href
  {\doibase 10.1126/science.1258480} {\bibfield  {journal} {\bibinfo  {journal}
  {Science}\ }\textbf {\bibinfo {volume} {346}},\ \bibinfo {pages} {975}
  (\bibinfo {year} {2014})}\BibitemShut {NoStop}%
\bibitem [{\citenamefont {Jing}\ \emph {et~al.}(2014)\citenamefont {Jing},
  \citenamefont {\"Ozdemir}, \citenamefont {L\"u}, \citenamefont {Zhang},
  \citenamefont {Yang},\ and\ \citenamefont {Nori}}]{PL1}%
  \BibitemOpen
  \bibfield  {author} {\bibinfo {author} {\bibfnamefont {H.}~\bibnamefont
  {Jing}}, \bibinfo {author} {\bibfnamefont {S.~K.}\ \bibnamefont {\"Ozdemir}},
  \bibinfo {author} {\bibfnamefont {X.-Y.}\ \bibnamefont {L\"u}}, \bibinfo
  {author} {\bibfnamefont {J.}~\bibnamefont {Zhang}}, \bibinfo {author}
  {\bibfnamefont {L.}~\bibnamefont {Yang}}, \ and\ \bibinfo {author}
  {\bibfnamefont {F.}~\bibnamefont {Nori}},\ }\href {\doibase
  10.1103/PhysRevLett.113.053604} {\bibfield  {journal} {\bibinfo  {journal}
  {Phys. Rev. Lett.}\ }\textbf {\bibinfo {volume} {113}},\ \bibinfo {pages}
  {053604} (\bibinfo {year} {2014})}\BibitemShut {NoStop}%
\bibitem [{\citenamefont {Feng}\ \emph {et~al.}(2014)\citenamefont {Feng},
  \citenamefont {Wong}, \citenamefont {Ma}, \citenamefont {Wang},\ and\
  \citenamefont {Zhang}}]{PL2}%
  \BibitemOpen
  \bibfield  {author} {\bibinfo {author} {\bibfnamefont {L.}~\bibnamefont
  {Feng}}, \bibinfo {author} {\bibfnamefont {Z.~J.}\ \bibnamefont {Wong}},
  \bibinfo {author} {\bibfnamefont {R.-M.}\ \bibnamefont {Ma}}, \bibinfo
  {author} {\bibfnamefont {Y.}~\bibnamefont {Wang}}, \ and\ \bibinfo {author}
  {\bibfnamefont {X.}~\bibnamefont {Zhang}},\ }\href {\doibase
  10.1126/science.1258479} {\bibfield  {journal} {\bibinfo  {journal}
  {Science}\ }\textbf {\bibinfo {volume} {346}},\ \bibinfo {pages} {972}
  (\bibinfo {year} {2014})}\BibitemShut {NoStop}%
\bibitem [{\citenamefont {Kulishov}\ \emph {et~al.}(2005)\citenamefont
  {Kulishov}, \citenamefont {Laniel}, \citenamefont {B\'{e}langer},
  \citenamefont {{n}a},\ and\ \citenamefont {Plant}}]{NR1}%
  \BibitemOpen
  \bibfield  {author} {\bibinfo {author} {\bibfnamefont {M.}~\bibnamefont
  {Kulishov}}, \bibinfo {author} {\bibfnamefont {J.~M.}\ \bibnamefont
  {Laniel}}, \bibinfo {author} {\bibfnamefont {N.}~\bibnamefont
  {B\'{e}langer}}, \bibinfo {author} {\bibfnamefont {J.~A.}\ \bibnamefont
  {{n}a}}, \ and\ \bibinfo {author} {\bibfnamefont {D.~V.}\ \bibnamefont
  {Plant}},\ }\href {\doibase 10.1364/OPEX.13.003068} {\bibfield  {journal}
  {\bibinfo  {journal} {Opt. Express}\ }\textbf {\bibinfo {volume} {13}},\
  \bibinfo {pages} {3068} (\bibinfo {year} {2005})}\BibitemShut {NoStop}%
\bibitem [{\citenamefont {Lin}\ \emph {et~al.}(2011)\citenamefont {Lin},
  \citenamefont {Ramezani}, \citenamefont {Eichelkraut}, \citenamefont
  {Kottos}, \citenamefont {Cao},\ and\ \citenamefont {Christodoulides}}]{UI1}%
  \BibitemOpen
  \bibfield  {author} {\bibinfo {author} {\bibfnamefont {Z.}~\bibnamefont
  {Lin}}, \bibinfo {author} {\bibfnamefont {H.}~\bibnamefont {Ramezani}},
  \bibinfo {author} {\bibfnamefont {T.}~\bibnamefont {Eichelkraut}}, \bibinfo
  {author} {\bibfnamefont {T.}~\bibnamefont {Kottos}}, \bibinfo {author}
  {\bibfnamefont {H.}~\bibnamefont {Cao}}, \ and\ \bibinfo {author}
  {\bibfnamefont {D.~N.}\ \bibnamefont {Christodoulides}},\ }\href {\doibase
  10.1103/PhysRevLett.106.213901} {\bibfield  {journal} {\bibinfo  {journal}
  {Phys. Rev. Lett.}\ }\textbf {\bibinfo {volume} {106}},\ \bibinfo {pages}
  {213901} (\bibinfo {year} {2011})}\BibitemShut {NoStop}%
\bibitem [{\citenamefont {Feng}\ \emph {et~al.}(2013)\citenamefont {Feng},
  \citenamefont {Xu}, \citenamefont {Fegadolli}, \citenamefont {Lu},
  \citenamefont {Oliveira}, \citenamefont {Almeida},\ and\ \citenamefont
  {Chen}}]{UI2}%
  \BibitemOpen
  \bibfield  {author} {\bibinfo {author} {\bibfnamefont {L.}~\bibnamefont
  {Feng}}, \bibinfo {author} {\bibfnamefont {Y.-L.}\ \bibnamefont {Xu}},
  \bibinfo {author} {\bibfnamefont {W.~S.}\ \bibnamefont {Fegadolli}}, \bibinfo
  {author} {\bibfnamefont {M.-H.}\ \bibnamefont {Lu}}, \bibinfo {author}
  {\bibfnamefont {J.~E.~B.}\ \bibnamefont {Oliveira}}, \bibinfo {author}
  {\bibfnamefont {V.~R.}\ \bibnamefont {Almeida}}, \ and\ \bibinfo {author}
  {\bibfnamefont {Y.-F.}\ \bibnamefont {Chen}},\ }\href {\doibase
  10.1038/nmat3495} {\bibfield  {journal} {\bibinfo  {journal} {Nature Mater.}\
  }\textbf {\bibinfo {volume} {12}},\ \bibinfo {pages} {108} (\bibinfo {year}
  {2013})}\BibitemShut {NoStop}%
\bibitem [{\citenamefont {Yin}\ and\ \citenamefont {Zhang}(2013)}]{UI3}%
  \BibitemOpen
  \bibfield  {author} {\bibinfo {author} {\bibfnamefont {X.}~\bibnamefont
  {Yin}}\ and\ \bibinfo {author} {\bibfnamefont {X.}~\bibnamefont {Zhang}},\
  }\href {\doibase 10.1038/nmat3576} {\bibfield  {journal} {\bibinfo  {journal}
  {Nature Mater.}\ }\textbf {\bibinfo {volume} {12}},\ \bibinfo {pages} {175}
  (\bibinfo {year} {2013})}\BibitemShut {NoStop}%
\bibitem [{\citenamefont {Rivolta}\ and\ \citenamefont {Maes}(2016)}]{UI4}%
  \BibitemOpen
  \bibfield  {author} {\bibinfo {author} {\bibfnamefont {N.~X.~A.}\
  \bibnamefont {Rivolta}}\ and\ \bibinfo {author} {\bibfnamefont
  {B.}~\bibnamefont {Maes}},\ }\href {\doibase 10.1103/PhysRevA.94.053854}
  {\bibfield  {journal} {\bibinfo  {journal} {Phys. Rev. A}\ }\textbf {\bibinfo
  {volume} {94}},\ \bibinfo {pages} {053854} (\bibinfo {year}
  {2016})}\BibitemShut {NoStop}%
\bibitem [{\citenamefont {Sarisaman}\ and\ \citenamefont {Tas}(2018)}]{UI5}%
  \BibitemOpen
  \bibfield  {author} {\bibinfo {author} {\bibfnamefont {M.}~\bibnamefont
  {Sarisaman}}\ and\ \bibinfo {author} {\bibfnamefont {M.}~\bibnamefont
  {Tas}},\ }\href {\doibase 10.1103/PhysRevB.97.045409} {\bibfield  {journal}
  {\bibinfo  {journal} {Phys. Rev. B}\ }\textbf {\bibinfo {volume} {97}},\
  \bibinfo {pages} {045409} (\bibinfo {year} {2018})}\BibitemShut {NoStop}%
\bibitem [{\citenamefont {Kre\ifmmode \check{s}\else
  \v{s}\fi{}i\ifmmode~\acute{c}\else \'{c}\fi{}}\ \emph
  {et~al.}(2022)\citenamefont {Kre\ifmmode \check{s}\else
  \v{s}\fi{}i\ifmmode~\acute{c}\else \'{c}\fi{}}, \citenamefont {Makris},
  \citenamefont {Leonhardt},\ and\ \citenamefont {Rotter}}]{UI6}%
  \BibitemOpen
  \bibfield  {author} {\bibinfo {author} {\bibfnamefont {I.}~\bibnamefont
  {Kre\ifmmode \check{s}\else \v{s}\fi{}i\ifmmode~\acute{c}\else \'{c}\fi{}}},
  \bibinfo {author} {\bibfnamefont {K.~G.}\ \bibnamefont {Makris}}, \bibinfo
  {author} {\bibfnamefont {U.}~\bibnamefont {Leonhardt}}, \ and\ \bibinfo
  {author} {\bibfnamefont {S.}~\bibnamefont {Rotter}},\ }\href {\doibase
  10.1103/PhysRevLett.128.183901} {\bibfield  {journal} {\bibinfo  {journal}
  {Phys. Rev. Lett.}\ }\textbf {\bibinfo {volume} {128}},\ \bibinfo {pages}
  {183901} (\bibinfo {year} {2022})}\BibitemShut {NoStop}%
\bibitem [{\citenamefont {Caucheteur}\ \emph {et~al.}(2015)\citenamefont
  {Caucheteur}, \citenamefont {Guo},\ and\ \citenamefont {Albert}}]{Grating1}%
  \BibitemOpen
  \bibfield  {author} {\bibinfo {author} {\bibfnamefont {C.}~\bibnamefont
  {Caucheteur}}, \bibinfo {author} {\bibfnamefont {T.}~\bibnamefont {Guo}}, \
  and\ \bibinfo {author} {\bibfnamefont {J.}~\bibnamefont {Albert}},\ }\href
  {\doibase 10.1007/s00216-014-8411-6} {\bibfield  {journal} {\bibinfo
  {journal} {Anal. Bioanal. Chem.}\ }\textbf {\bibinfo {volume} {407}},\
  \bibinfo {pages} {3883} (\bibinfo {year} {2015})}\BibitemShut {NoStop}%
\bibitem [{\citenamefont {Mihailov}(2012)}]{Grating2}%
  \BibitemOpen
  \bibfield  {author} {\bibinfo {author} {\bibfnamefont {S.~J.}\ \bibnamefont
  {Mihailov}},\ }\href {\doibase 10.3390/s120201898} {\bibfield  {journal}
  {\bibinfo  {journal} {Sensors}\ }\textbf {\bibinfo {volume} {12}},\ \bibinfo
  {pages} {1898} (\bibinfo {year} {2012})}\BibitemShut {NoStop}%
\bibitem [{\citenamefont {Zanutta}\ \emph {et~al.}(2014)\citenamefont
  {Zanutta}, \citenamefont {Bianco}, \citenamefont {Insausti},\ and\
  \citenamefont {Garz\'{o}n}}]{Grating3}%
  \BibitemOpen
  \bibfield  {author} {\bibinfo {author} {\bibfnamefont {A.}~\bibnamefont
  {Zanutta}}, \bibinfo {author} {\bibfnamefont {A.}~\bibnamefont {Bianco}},
  \bibinfo {author} {\bibfnamefont {M.}~\bibnamefont {Insausti}}, \ and\
  \bibinfo {author} {\bibfnamefont {F.}~\bibnamefont {Garz\'{o}n}},\ }in\ \href
  {\doibase 10.1117/12.2055743} {\emph {\bibinfo {booktitle} {Advances in
  Optical and Mechanical Technologies for Telescopes and Instrumentation}}},\
  Vol.\ \bibinfo {volume} {9151},\ \bibinfo {editor} {edited by\ \bibinfo
  {editor} {\bibfnamefont {R.}~\bibnamefont {Navarro}}, \bibinfo {editor}
  {\bibfnamefont {C.~R.}\ \bibnamefont {Cunningham}}, \ and\ \bibinfo {editor}
  {\bibfnamefont {A.~A.}\ \bibnamefont {Barto}}},\ \bibinfo {organization}
  {International Society for Optics and Photonics}\ (\bibinfo  {publisher}
  {SPIE},\ \bibinfo {year} {2014})\ pp.\ \bibinfo {pages} {1992 --
  2006}\BibitemShut {NoStop}%
\bibitem [{\citenamefont {Raijmakers}\ \emph {et~al.}(2019)\citenamefont
  {Raijmakers}, \citenamefont {Danilov}, \citenamefont {Eichel},\ and\
  \citenamefont {Notten}}]{Grating4}%
  \BibitemOpen
  \bibfield  {author} {\bibinfo {author} {\bibfnamefont {L.}~\bibnamefont
  {Raijmakers}}, \bibinfo {author} {\bibfnamefont {D.}~\bibnamefont {Danilov}},
  \bibinfo {author} {\bibfnamefont {R.-A.}\ \bibnamefont {Eichel}}, \ and\
  \bibinfo {author} {\bibfnamefont {P.}~\bibnamefont {Notten}},\ }\href
  {\doibase https://doi.org/10.1016/j.apenergy.2019.02.078} {\bibfield
  {journal} {\bibinfo  {journal} {Appl. Energy}\ }\textbf {\bibinfo {volume}
  {240}},\ \bibinfo {pages} {918} (\bibinfo {year} {2019})}\BibitemShut
  {NoStop}%
\bibitem [{\citenamefont {Ling}\ \emph {et~al.}(1998)\citenamefont {Ling},
  \citenamefont {Li},\ and\ \citenamefont {Xiao}}]{EIG1}%
  \BibitemOpen
  \bibfield  {author} {\bibinfo {author} {\bibfnamefont {H.~Y.}\ \bibnamefont
  {Ling}}, \bibinfo {author} {\bibfnamefont {Y.-Q.}\ \bibnamefont {Li}}, \ and\
  \bibinfo {author} {\bibfnamefont {M.}~\bibnamefont {Xiao}},\ }\href {\doibase
  10.1103/PhysRevA.57.1338} {\bibfield  {journal} {\bibinfo  {journal} {Phys.
  Rev. A}\ }\textbf {\bibinfo {volume} {57}},\ \bibinfo {pages} {1338}
  (\bibinfo {year} {1998})}\BibitemShut {NoStop}%
\bibitem [{\citenamefont {Mitsunaga}\ and\ \citenamefont {Imoto}(1999)}]{EIG2}%
  \BibitemOpen
  \bibfield  {author} {\bibinfo {author} {\bibfnamefont {M.}~\bibnamefont
  {Mitsunaga}}\ and\ \bibinfo {author} {\bibfnamefont {N.}~\bibnamefont
  {Imoto}},\ }\href {\doibase 10.1103/PhysRevA.59.4773} {\bibfield  {journal}
  {\bibinfo  {journal} {Phys. Rev. A}\ }\textbf {\bibinfo {volume} {59}},\
  \bibinfo {pages} {4773} (\bibinfo {year} {1999})}\BibitemShut {NoStop}%
\bibitem [{\citenamefont {Cardoso}\ and\ \citenamefont {Tabosa}(2002)}]{EIG3}%
  \BibitemOpen
  \bibfield  {author} {\bibinfo {author} {\bibfnamefont {G.~C.}\ \bibnamefont
  {Cardoso}}\ and\ \bibinfo {author} {\bibfnamefont {J.~W.~R.}\ \bibnamefont
  {Tabosa}},\ }\href {\doibase 10.1103/PhysRevA.65.033803} {\bibfield
  {journal} {\bibinfo  {journal} {Phys. Rev. A}\ }\textbf {\bibinfo {volume}
  {65}},\ \bibinfo {pages} {033803} (\bibinfo {year} {2002})}\BibitemShut
  {NoStop}%
\bibitem [{\citenamefont {Fleischhauer}\ \emph {et~al.}(2005)\citenamefont
  {Fleischhauer}, \citenamefont {Imamoglu},\ and\ \citenamefont
  {Marangos}}]{EIT-review}%
  \BibitemOpen
  \bibfield  {author} {\bibinfo {author} {\bibfnamefont {M.}~\bibnamefont
  {Fleischhauer}}, \bibinfo {author} {\bibfnamefont {A.}~\bibnamefont
  {Imamoglu}}, \ and\ \bibinfo {author} {\bibfnamefont {J.~P.}\ \bibnamefont
  {Marangos}},\ }\href {\doibase 10.1103/RevModPhys.77.633} {\bibfield
  {journal} {\bibinfo  {journal} {Rev. Mod. Phys.}\ }\textbf {\bibinfo {volume}
  {77}},\ \bibinfo {pages} {633} (\bibinfo {year} {2005})}\BibitemShut
  {NoStop}%
\bibitem [{\citenamefont {Zhao}\ \emph {et~al.}(2010)\citenamefont {Zhao},
  \citenamefont {Duan},\ and\ \citenamefont {Yelin}}]{EIG4}%
  \BibitemOpen
  \bibfield  {author} {\bibinfo {author} {\bibfnamefont {L.}~\bibnamefont
  {Zhao}}, \bibinfo {author} {\bibfnamefont {W.}~\bibnamefont {Duan}}, \ and\
  \bibinfo {author} {\bibfnamefont {S.~F.}\ \bibnamefont {Yelin}},\ }\href
  {\doibase 10.1103/PhysRevA.82.013809} {\bibfield  {journal} {\bibinfo
  {journal} {Phys. Rev. A}\ }\textbf {\bibinfo {volume} {82}},\ \bibinfo
  {pages} {013809} (\bibinfo {year} {2010})}\BibitemShut {NoStop}%
\bibitem [{\citenamefont {de~Araujo}(2010)}]{EIGp1}%
  \BibitemOpen
  \bibfield  {author} {\bibinfo {author} {\bibfnamefont {L.~E.~E.}\
  \bibnamefont {de~Araujo}},\ }\href {\doibase 10.1364/OL.35.000977} {\bibfield
   {journal} {\bibinfo  {journal} {Opt. Lett.}\ }\textbf {\bibinfo {volume}
  {35}},\ \bibinfo {pages} {977} (\bibinfo {year} {2010})}\BibitemShut
  {NoStop}%
\bibitem [{\citenamefont {Carvalho}\ and\ \citenamefont
  {de~Araujo}(2011)}]{EIGp2}%
  \BibitemOpen
  \bibfield  {author} {\bibinfo {author} {\bibfnamefont {S.~A.}\ \bibnamefont
  {Carvalho}}\ and\ \bibinfo {author} {\bibfnamefont {L.~E.~E.}\ \bibnamefont
  {de~Araujo}},\ }\href {\doibase 10.1364/OE.19.001936} {\bibfield  {journal}
  {\bibinfo  {journal} {Opt. Express}\ }\textbf {\bibinfo {volume} {19}},\
  \bibinfo {pages} {1936} (\bibinfo {year} {2011})}\BibitemShut {NoStop}%
\bibitem [{\citenamefont {Vafafard}\ and\ \citenamefont
  {Mahmoudi}(2015)}]{EIGp3}%
  \BibitemOpen
  \bibfield  {author} {\bibinfo {author} {\bibfnamefont {A.}~\bibnamefont
  {Vafafard}}\ and\ \bibinfo {author} {\bibfnamefont {M.}~\bibnamefont
  {Mahmoudi}},\ }\href {\doibase 10.1364/AO.54.010613} {\bibfield  {journal}
  {\bibinfo  {journal} {Appl. Opt.}\ }\textbf {\bibinfo {volume} {54}},\
  \bibinfo {pages} {10613} (\bibinfo {year} {2015})}\BibitemShut {NoStop}%
\bibitem [{\citenamefont {Liu}\ \emph {et~al.}(2016)\citenamefont {Liu},
  \citenamefont {Tian}, \citenamefont {Wang}, \citenamefont {Yan},\ and\
  \citenamefont {Wu}}]{RydbergEIG1}%
  \BibitemOpen
  \bibfield  {author} {\bibinfo {author} {\bibfnamefont {Y.-M.}\ \bibnamefont
  {Liu}}, \bibinfo {author} {\bibfnamefont {X.-D.}\ \bibnamefont {Tian}},
  \bibinfo {author} {\bibfnamefont {X.}~\bibnamefont {Wang}}, \bibinfo {author}
  {\bibfnamefont {D.}~\bibnamefont {Yan}}, \ and\ \bibinfo {author}
  {\bibfnamefont {J.-H.}\ \bibnamefont {Wu}},\ }\href {\doibase
  10.1364/OL.41.000408} {\bibfield  {journal} {\bibinfo  {journal} {Opt.
  Lett.}\ }\textbf {\bibinfo {volume} {41}},\ \bibinfo {pages} {408} (\bibinfo
  {year} {2016})}\BibitemShut {NoStop}%
\bibitem [{\citenamefont {Asghar}\ \emph {et~al.}(2016)\citenamefont {Asghar},
  \citenamefont {Ziauddin}, \citenamefont {Qamar},\ and\ \citenamefont
  {Qamar}}]{RydbergEIG2}%
  \BibitemOpen
  \bibfield  {author} {\bibinfo {author} {\bibfnamefont {S.}~\bibnamefont
  {Asghar}}, \bibinfo {author} {\bibnamefont {Ziauddin}}, \bibinfo {author}
  {\bibfnamefont {S.}~\bibnamefont {Qamar}}, \ and\ \bibinfo {author}
  {\bibfnamefont {S.}~\bibnamefont {Qamar}},\ }\href {\doibase
  10.1103/PhysRevA.94.033823} {\bibfield  {journal} {\bibinfo  {journal} {Phys.
  Rev. A}\ }\textbf {\bibinfo {volume} {94}},\ \bibinfo {pages} {033823}
  (\bibinfo {year} {2016})}\BibitemShut {NoStop}%
\bibitem [{\citenamefont {Zhu}\ \emph {et~al.}(2016)\citenamefont {Zhu},
  \citenamefont {Xu}, \citenamefont {Zou}, \citenamefont {Sun}, \citenamefont
  {He}, \citenamefont {Lu}, \citenamefont {Liu},\ and\ \citenamefont
  {Chen}}]{PTG1}%
  \BibitemOpen
  \bibfield  {author} {\bibinfo {author} {\bibfnamefont {X.-Y.}\ \bibnamefont
  {Zhu}}, \bibinfo {author} {\bibfnamefont {Y.-L.}\ \bibnamefont {Xu}},
  \bibinfo {author} {\bibfnamefont {Y.}~\bibnamefont {Zou}}, \bibinfo {author}
  {\bibfnamefont {X.-C.}\ \bibnamefont {Sun}}, \bibinfo {author} {\bibfnamefont
  {C.}~\bibnamefont {He}}, \bibinfo {author} {\bibfnamefont {M.-H.}\
  \bibnamefont {Lu}}, \bibinfo {author} {\bibfnamefont {X.-P.}\ \bibnamefont
  {Liu}}, \ and\ \bibinfo {author} {\bibfnamefont {Y.-F.}\ \bibnamefont
  {Chen}},\ }\href {\doibase 10.1063/1.4962639} {\bibfield  {journal} {\bibinfo
   {journal} {Appl. Phys. Lett.}\ }\textbf {\bibinfo {volume} {109}},\ \bibinfo
  {pages} {111101} (\bibinfo {year} {2016})}\BibitemShut {NoStop}%
\bibitem [{\citenamefont {Liu}\ \emph {et~al.}(2017)\citenamefont {Liu},
  \citenamefont {Gao}, \citenamefont {Fan},\ and\ \citenamefont {Wu}}]{PTG2}%
  \BibitemOpen
  \bibfield  {author} {\bibinfo {author} {\bibfnamefont {Y.-M.}\ \bibnamefont
  {Liu}}, \bibinfo {author} {\bibfnamefont {F.}~\bibnamefont {Gao}}, \bibinfo
  {author} {\bibfnamefont {C.-H.}\ \bibnamefont {Fan}}, \ and\ \bibinfo
  {author} {\bibfnamefont {J.-H.}\ \bibnamefont {Wu}},\ }\href {\doibase
  10.1364/OL.42.004283} {\bibfield  {journal} {\bibinfo  {journal} {Opt.
  Lett.}\ }\textbf {\bibinfo {volume} {42}},\ \bibinfo {pages} {4283} (\bibinfo
  {year} {2017})}\BibitemShut {NoStop}%
\bibitem [{\citenamefont {Bushuev}\ \emph {et~al.}(2017)\citenamefont
  {Bushuev}, \citenamefont {Dergacheva},\ and\ \citenamefont
  {Mantsyzov}}]{PTG3}%
  \BibitemOpen
  \bibfield  {author} {\bibinfo {author} {\bibfnamefont {V.~A.}\ \bibnamefont
  {Bushuev}}, \bibinfo {author} {\bibfnamefont {L.~V.}\ \bibnamefont
  {Dergacheva}}, \ and\ \bibinfo {author} {\bibfnamefont {B.~I.}\ \bibnamefont
  {Mantsyzov}},\ }\href {\doibase 10.1103/PhysRevA.95.033843} {\bibfield
  {journal} {\bibinfo  {journal} {Phys. Rev. A}\ }\textbf {\bibinfo {volume}
  {95}},\ \bibinfo {pages} {033843} (\bibinfo {year} {2017})}\BibitemShut
  {NoStop}%
\bibitem [{\citenamefont {Zhang}\ \emph {et~al.}(2018)\citenamefont {Zhang},
  \citenamefont {Yang}, \citenamefont {Feng}, \citenamefont {Sheng},
  \citenamefont {Zhang}, \citenamefont {Zhang},\ and\ \citenamefont
  {Xiao}}]{PTG4}%
  \BibitemOpen
  \bibfield  {author} {\bibinfo {author} {\bibfnamefont {Z.}~\bibnamefont
  {Zhang}}, \bibinfo {author} {\bibfnamefont {L.}~\bibnamefont {Yang}},
  \bibinfo {author} {\bibfnamefont {J.}~\bibnamefont {Feng}}, \bibinfo {author}
  {\bibfnamefont {J.}~\bibnamefont {Sheng}}, \bibinfo {author} {\bibfnamefont
  {Y.}~\bibnamefont {Zhang}}, \bibinfo {author} {\bibfnamefont
  {Y.}~\bibnamefont {Zhang}}, \ and\ \bibinfo {author} {\bibfnamefont
  {M.}~\bibnamefont {Xiao}},\ }\href {\doibase
  https://doi.org/10.1002/lpor.201800155} {\bibfield  {journal} {\bibinfo
  {journal} {Laser Photonics Rev.}\ }\textbf {\bibinfo {volume} {12}},\
  \bibinfo {pages} {1800155} (\bibinfo {year} {2018})}\BibitemShut {NoStop}%
\bibitem [{\citenamefont {Shui}\ \emph {et~al.}(2018)\citenamefont {Shui},
  \citenamefont {Yang}, \citenamefont {Liu}, \citenamefont {Li},\ and\
  \citenamefont {Zhu}}]{PTG5}%
  \BibitemOpen
  \bibfield  {author} {\bibinfo {author} {\bibfnamefont {T.}~\bibnamefont
  {Shui}}, \bibinfo {author} {\bibfnamefont {W.-X.}\ \bibnamefont {Yang}},
  \bibinfo {author} {\bibfnamefont {S.}~\bibnamefont {Liu}}, \bibinfo {author}
  {\bibfnamefont {L.}~\bibnamefont {Li}}, \ and\ \bibinfo {author}
  {\bibfnamefont {Z.}~\bibnamefont {Zhu}},\ }\href {\doibase
  10.1103/PhysRevA.97.033819} {\bibfield  {journal} {\bibinfo  {journal} {Phys.
  Rev. A}\ }\textbf {\bibinfo {volume} {97}},\ \bibinfo {pages} {033819}
  (\bibinfo {year} {2018})}\BibitemShut {NoStop}%
\bibitem [{\citenamefont {Ma}\ \emph {et~al.}(2019)\citenamefont {Ma},
  \citenamefont {Yu}, \citenamefont {Zhao},\ and\ \citenamefont {Qian}}]{PTG6}%
  \BibitemOpen
  \bibfield  {author} {\bibinfo {author} {\bibfnamefont {D.}~\bibnamefont
  {Ma}}, \bibinfo {author} {\bibfnamefont {D.}~\bibnamefont {Yu}}, \bibinfo
  {author} {\bibfnamefont {X.-D.}\ \bibnamefont {Zhao}}, \ and\ \bibinfo
  {author} {\bibfnamefont {J.}~\bibnamefont {Qian}},\ }\href {\doibase
  10.1103/PhysRevA.99.033826} {\bibfield  {journal} {\bibinfo  {journal} {Phys.
  Rev. A}\ }\textbf {\bibinfo {volume} {99}},\ \bibinfo {pages} {033826}
  (\bibinfo {year} {2019})}\BibitemShut {NoStop}%
\bibitem [{\citenamefont {Liu}\ \emph {et~al.}(2019)\citenamefont {Liu},
  \citenamefont {Gao}, \citenamefont {Wu}, \citenamefont {Artoni},\ and\
  \citenamefont {La~Rocca}}]{PTG7}%
  \BibitemOpen
  \bibfield  {author} {\bibinfo {author} {\bibfnamefont {Y.-M.}\ \bibnamefont
  {Liu}}, \bibinfo {author} {\bibfnamefont {F.}~\bibnamefont {Gao}}, \bibinfo
  {author} {\bibfnamefont {J.-H.}\ \bibnamefont {Wu}}, \bibinfo {author}
  {\bibfnamefont {M.}~\bibnamefont {Artoni}}, \ and\ \bibinfo {author}
  {\bibfnamefont {G.~C.}\ \bibnamefont {La~Rocca}},\ }\href {\doibase
  10.1103/PhysRevA.100.043801} {\bibfield  {journal} {\bibinfo  {journal}
  {Phys. Rev. A}\ }\textbf {\bibinfo {volume} {100}},\ \bibinfo {pages}
  {043801} (\bibinfo {year} {2019})}\BibitemShut {NoStop}%
\bibitem [{\citenamefont {Hang}\ \emph {et~al.}(2019)\citenamefont {Hang},
  \citenamefont {Li},\ and\ \citenamefont {Huang}}]{PTG8}%
  \BibitemOpen
  \bibfield  {author} {\bibinfo {author} {\bibfnamefont {C.}~\bibnamefont
  {Hang}}, \bibinfo {author} {\bibfnamefont {W.}~\bibnamefont {Li}}, \ and\
  \bibinfo {author} {\bibfnamefont {G.}~\bibnamefont {Huang}},\ }\href
  {\doibase 10.1103/PhysRevA.100.043807} {\bibfield  {journal} {\bibinfo
  {journal} {Phys. Rev. A}\ }\textbf {\bibinfo {volume} {100}},\ \bibinfo
  {pages} {043807} (\bibinfo {year} {2019})}\BibitemShut {NoStop}%
\bibitem [{\citenamefont {Yang}\ \emph {et~al.}(2019)\citenamefont {Yang},
  \citenamefont {Jia}, \citenamefont {Bi}, \citenamefont {Zhao},\ and\
  \citenamefont {Yang}}]{AA1}%
  \BibitemOpen
  \bibfield  {author} {\bibinfo {author} {\bibfnamefont {Y.}~\bibnamefont
  {Yang}}, \bibinfo {author} {\bibfnamefont {H.}~\bibnamefont {Jia}}, \bibinfo
  {author} {\bibfnamefont {Y.}~\bibnamefont {Bi}}, \bibinfo {author}
  {\bibfnamefont {H.}~\bibnamefont {Zhao}}, \ and\ \bibinfo {author}
  {\bibfnamefont {J.}~\bibnamefont {Yang}},\ }\href {\doibase
  10.1103/PhysRevApplied.12.034040} {\bibfield  {journal} {\bibinfo  {journal}
  {Phys. Rev. Applied}\ }\textbf {\bibinfo {volume} {12}},\ \bibinfo {pages}
  {034040} (\bibinfo {year} {2019})}\BibitemShut {NoStop}%
\bibitem [{\citenamefont {Miyake}\ and\ \citenamefont {Uyeda}(1950)}]{FL1}%
  \BibitemOpen
  \bibfield  {author} {\bibinfo {author} {\bibfnamefont {S.}~\bibnamefont
  {Miyake}}\ and\ \bibinfo {author} {\bibfnamefont {R.}~\bibnamefont {Uyeda}},\
  }\href {\doibase 10.1107/S0365110X5000080X} {\bibfield  {journal} {\bibinfo
  {journal} {Acta. Cryst.}\ }\textbf {\bibinfo {volume} {3}},\ \bibinfo {pages}
  {314} (\bibinfo {year} {1950})}\BibitemShut {NoStop}%
\bibitem [{\citenamefont {Miyake}\ and\ \citenamefont {Uyeda}(1955)}]{FL2}%
  \BibitemOpen
  \bibfield  {author} {\bibinfo {author} {\bibfnamefont {S.}~\bibnamefont
  {Miyake}}\ and\ \bibinfo {author} {\bibfnamefont {R.}~\bibnamefont {Uyeda}},\
  }\href {\doibase 10.1107/S0365110X55001023} {\bibfield  {journal} {\bibinfo
  {journal} {Acta. Cryst.}\ }\textbf {\bibinfo {volume} {8}},\ \bibinfo {pages}
  {335} (\bibinfo {year} {1955})}\BibitemShut {NoStop}%
\bibitem [{\citenamefont {Deb}\ \emph {et~al.}(2020)\citenamefont {Deb},
  \citenamefont {Cao}, \citenamefont {Han}, \citenamefont {Holtz},
  \citenamefont {Xie}, \citenamefont {Park}, \citenamefont {Hovden},\ and\
  \citenamefont {Muller}}]{FL3}%
  \BibitemOpen
  \bibfield  {author} {\bibinfo {author} {\bibfnamefont {P.}~\bibnamefont
  {Deb}}, \bibinfo {author} {\bibfnamefont {M.~C.}\ \bibnamefont {Cao}},
  \bibinfo {author} {\bibfnamefont {Y.}~\bibnamefont {Han}}, \bibinfo {author}
  {\bibfnamefont {M.~E.}\ \bibnamefont {Holtz}}, \bibinfo {author}
  {\bibfnamefont {S.}~\bibnamefont {Xie}}, \bibinfo {author} {\bibfnamefont
  {J.}~\bibnamefont {Park}}, \bibinfo {author} {\bibfnamefont {R.}~\bibnamefont
  {Hovden}}, \ and\ \bibinfo {author} {\bibfnamefont {D.~A.}\ \bibnamefont
  {Muller}},\ }\href {\doibase https://doi.org/10.1016/j.ultramic.2020.113019}
  {\bibfield  {journal} {\bibinfo  {journal} {Ultramicroscopy}\ }\textbf
  {\bibinfo {volume} {215}},\ \bibinfo {pages} {113019} (\bibinfo {year}
  {2020})}\BibitemShut {NoStop}%
\bibitem [{\citenamefont {Horsley}\ \emph {et~al.}(2015)\citenamefont
  {Horsley}, \citenamefont {Artoni},\ and\ \citenamefont {La~Rocca}}]{SKK1}%
  \BibitemOpen
  \bibfield  {author} {\bibinfo {author} {\bibfnamefont {S.~A.~R.}\
  \bibnamefont {Horsley}}, \bibinfo {author} {\bibfnamefont {M.}~\bibnamefont
  {Artoni}}, \ and\ \bibinfo {author} {\bibfnamefont {G.~C.}\ \bibnamefont
  {La~Rocca}},\ }\href {\doibase 10.1038/nphoton.2015.106} {\bibfield
  {journal} {\bibinfo  {journal} {Nature Photon.}\ }\textbf {\bibinfo {volume}
  {9}},\ \bibinfo {pages} {436} (\bibinfo {year} {2015})}\BibitemShut {NoStop}%
\bibitem [{\citenamefont {Ye}\ \emph {et~al.}(2017)\citenamefont {Ye},
  \citenamefont {Cao}, \citenamefont {Zhou}, \citenamefont {Huangfu},
  \citenamefont {Zheng},\ and\ \citenamefont {Ran}}]{SKK2}%
  \BibitemOpen
  \bibfield  {author} {\bibinfo {author} {\bibfnamefont {D.}~\bibnamefont
  {Ye}}, \bibinfo {author} {\bibfnamefont {C.}~\bibnamefont {Cao}}, \bibinfo
  {author} {\bibfnamefont {T.}~\bibnamefont {Zhou}}, \bibinfo {author}
  {\bibfnamefont {J.}~\bibnamefont {Huangfu}}, \bibinfo {author} {\bibfnamefont
  {G.}~\bibnamefont {Zheng}}, \ and\ \bibinfo {author} {\bibfnamefont
  {L.}~\bibnamefont {Ran}},\ }\href {\doibase 10.1038/s41467-017-00123-4}
  {\bibfield  {journal} {\bibinfo  {journal} {Nat. Commun.}\ }\textbf {\bibinfo
  {volume} {8}},\ \bibinfo {pages} {51} (\bibinfo {year} {2017})}\BibitemShut
  {NoStop}%
\bibitem [{\citenamefont {Baek}\ and\ \citenamefont {Park}(2017)}]{SKK3}%
  \BibitemOpen
  \bibfield  {author} {\bibinfo {author} {\bibfnamefont {Y.}~\bibnamefont
  {Baek}}\ and\ \bibinfo {author} {\bibfnamefont {Y.}~\bibnamefont {Park}},\
  }\href {\doibase 10.1038/s41566-021-00760-8} {\bibfield  {journal} {\bibinfo
  {journal} {Nature Photon.}\ }\textbf {\bibinfo {volume} {15}},\ \bibinfo
  {pages} {6} (\bibinfo {year} {2017})}\BibitemShut {NoStop}%
\bibitem [{\citenamefont {Zhang}\ \emph {et~al.}(2021)\citenamefont {Zhang},
  \citenamefont {Wu}, \citenamefont {Artoni},\ and\ \citenamefont
  {Rocca}}]{SKK4}%
  \BibitemOpen
  \bibfield  {author} {\bibinfo {author} {\bibfnamefont {Y.}~\bibnamefont
  {Zhang}}, \bibinfo {author} {\bibfnamefont {J.-H.}\ \bibnamefont {Wu}},
  \bibinfo {author} {\bibfnamefont {M.}~\bibnamefont {Artoni}}, \ and\ \bibinfo
  {author} {\bibfnamefont {G.~C.~L.}\ \bibnamefont {Rocca}},\ }\href {\doibase
  10.1364/OE.415879} {\bibfield  {journal} {\bibinfo  {journal} {Opt. Express}\
  }\textbf {\bibinfo {volume} {29}},\ \bibinfo {pages} {5890} (\bibinfo {year}
  {2021})}\BibitemShut {NoStop}%
\bibitem [{\citenamefont {Hua}\ \emph {et~al.}(2022)\citenamefont {Hua},
  \citenamefont {Liu}, \citenamefont {Lio}, \citenamefont {Zhang},
  \citenamefont {Wu}, \citenamefont {Artoni},\ and\ \citenamefont
  {La~Rocca}}]{HS}%
  \BibitemOpen
  \bibfield  {author} {\bibinfo {author} {\bibfnamefont {S.}~\bibnamefont
  {Hua}}, \bibinfo {author} {\bibfnamefont {Y.-M.}\ \bibnamefont {Liu}},
  \bibinfo {author} {\bibfnamefont {G.~E.}\ \bibnamefont {Lio}}, \bibinfo
  {author} {\bibfnamefont {X.-J.}\ \bibnamefont {Zhang}}, \bibinfo {author}
  {\bibfnamefont {J.-H.}\ \bibnamefont {Wu}}, \bibinfo {author} {\bibfnamefont
  {M.}~\bibnamefont {Artoni}}, \ and\ \bibinfo {author} {\bibfnamefont {G.~C.}\
  \bibnamefont {La~Rocca}},\ }\href {\doibase 10.1103/PhysRevResearch.4.023113}
  {\bibfield  {journal} {\bibinfo  {journal} {Phys. Rev. Research}\ }\textbf
  {\bibinfo {volume} {4}},\ \bibinfo {pages} {023113} (\bibinfo {year}
  {2022})}\BibitemShut {NoStop}%
\bibitem [{\citenamefont {Dammann}\ and\ \citenamefont
  {Görtler}(1971)}]{DAMMANN_1}%
  \BibitemOpen
  \bibfield  {author} {\bibinfo {author} {\bibfnamefont {H.}~\bibnamefont
  {Dammann}}\ and\ \bibinfo {author} {\bibfnamefont {K.}~\bibnamefont
  {Görtler}},\ }\href {\doibase https://doi.org/10.1016/0030-4018(71)90095-2}
  {\bibfield  {journal} {\bibinfo  {journal} {Opt. Commun.}\ }\textbf {\bibinfo
  {volume} {3}},\ \bibinfo {pages} {312} (\bibinfo {year} {1971})}\BibitemShut
  {NoStop}%
\bibitem [{\citenamefont {Dammann}\ and\ \citenamefont
  {Klotz}(1977)}]{DAMMANN_2}%
  \BibitemOpen
  \bibfield  {author} {\bibinfo {author} {\bibfnamefont {H.}~\bibnamefont
  {Dammann}}\ and\ \bibinfo {author} {\bibfnamefont {E.}~\bibnamefont
  {Klotz}},\ }\href {\doibase 10.1080/713819570} {\bibfield  {journal}
  {\bibinfo  {journal} {Opt. Acta: I. J. Opt.}\ }\textbf {\bibinfo {volume}
  {24}},\ \bibinfo {pages} {505} (\bibinfo {year} {1977})}\BibitemShut
  {NoStop}%
\bibitem [{\citenamefont {Turunen}\ \emph {et~al.}(1989)\citenamefont
  {Turunen}, \citenamefont {Vasara}, \citenamefont {Westerholm},\ and\
  \citenamefont {Salin}}]{DAMMANN_3}%
  \BibitemOpen
  \bibfield  {author} {\bibinfo {author} {\bibfnamefont {J.}~\bibnamefont
  {Turunen}}, \bibinfo {author} {\bibfnamefont {A.}~\bibnamefont {Vasara}},
  \bibinfo {author} {\bibfnamefont {J.}~\bibnamefont {Westerholm}}, \ and\
  \bibinfo {author} {\bibfnamefont {A.}~\bibnamefont {Salin}},\ }\href
  {\doibase https://doi.org/10.1016/0030-4018(89)90358-1} {\bibfield  {journal}
  {\bibinfo  {journal} {Opt. Commun.}\ }\textbf {\bibinfo {volume} {74}},\
  \bibinfo {pages} {245} (\bibinfo {year} {1989})}\BibitemShut {NoStop}%
\bibitem [{\citenamefont {Zhou}\ and\ \citenamefont {Liu}(1995)}]{DAMMANN_4}%
  \BibitemOpen
  \bibfield  {author} {\bibinfo {author} {\bibfnamefont {C.}~\bibnamefont
  {Zhou}}\ and\ \bibinfo {author} {\bibfnamefont {L.}~\bibnamefont {Liu}},\
  }\href {\doibase 10.1364/AO.34.005961} {\bibfield  {journal} {\bibinfo
  {journal} {Appl. Opt.}\ }\textbf {\bibinfo {volume} {34}},\ \bibinfo {pages}
  {5961} (\bibinfo {year} {1995})}\BibitemShut {NoStop}%
\bibitem [{\citenamefont {Pang}\ \emph {et~al.}(2019)\citenamefont {Pang},
  \citenamefont {Cao}, \citenamefont {Liu}, \citenamefont {Shi},\ and\
  \citenamefont {Deng}}]{DAMMANN_5}%
  \BibitemOpen
  \bibfield  {author} {\bibinfo {author} {\bibfnamefont {H.}~\bibnamefont
  {Pang}}, \bibinfo {author} {\bibfnamefont {A.}~\bibnamefont {Cao}}, \bibinfo
  {author} {\bibfnamefont {W.}~\bibnamefont {Liu}}, \bibinfo {author}
  {\bibfnamefont {L.}~\bibnamefont {Shi}}, \ and\ \bibinfo {author}
  {\bibfnamefont {Q.}~\bibnamefont {Deng}},\ }\href {\doibase
  10.1109/JPHOT.2019.2899903} {\bibfield  {journal} {\bibinfo  {journal} {IEEE
  Photon. J.}\ }\textbf {\bibinfo {volume} {11}},\ \bibinfo {pages} {1}
  (\bibinfo {year} {2019})}\BibitemShut {NoStop}%
\end{thebibliography}%

\end{document}